\documentclass [12pt,a4paper] {article}
\usepackage{graphicx}
\usepackage{epsfig}

\usepackage[symbol]{footmisc}

\begin {document}
\begin{center}

{\large{\bf Multistrange Hyperon Production on Nuclear Targets}}
\vspace{.5cm}

G.H.~Arakelyan$^1$$\dagger$, C.~Merino$^2$, and Yu.M.~Shabelski$^3$\\
\vspace{.5cm}

$^1$A.Alikhanyan National Scientific Laboratory
(Yerevan Physics Institute)\\
Yerevan, 0036, Armenia\\
E-mail: argev@mail.yerphi.am

\vspace{.2cm}

$^2$Departamento de F\'\i sica de Part\'\i culas, Facultade de F\'\i sica \\
Instituto Galego de F\'\i sica de Altas Enerx\'\i as (IGFAE) \\
Universidade de Santiago de Compostela, Galiza, Spain \\
E-mail: carlos.merino@usc.es

\vspace{.2cm}

$^3$Petersburg Nuclear Physics Institute \\
NCR Kurchatov Institute \\
Gatchina, St.Petersburg 188300 Russia \\
E-mail: shabelsk@thd.pnpi.spb.ru
\vskip 0.5 cm

\end{center}

\begin{abstract}
We consider the experimental data on yields of protons, strange $\Lambda^{\prime}$s, and multistrange  
baryons ($\Xi$, $\Omega$), and antibaryons production on nuclear targets, and the experimental ratios of 
multistrange to strange antibaryon production, at the energy region from SPS up to LHC, and compare 
them to the results of the Quark-Gluon String Model calculations.
In the case of heavy nucleus collisions, the experimental dependence of the $\overline{\Xi}^+/\overline{\Lambda}$,
and, in particular, of the $\overline{\Omega}^+/\overline{\Lambda}$ ratios, on the centrality of the collision,
shows a manifest violation of quark combinatorial rules.
\end{abstract}

\vskip 1cm


\footnote[0]{$\dagger$ Deceased}

\newpage
{
\section{Introduction}

The Quark-Gluon String Model (QGSM)~\cite{KTM82,KTM,Kaid,K20} is based on the Dual Topological Unitarization (DTU), 
Regge phenomenology~\cite{RV}, and nonperturbative notions of QCD~\cite{GMV}.
 
In QGSM, high energy interactions are considered as proceeding via the exchange of one or several Pomerons.
The cut of at least some of those Pomerons determines the inelastic scattering amplitude of the particle
production processes, that occur trough the production and subsequent decay of the quark-gluon strings resulting
of the cut of the Pomerons.

In the case of interaction with a nuclear target, the Multiple Scattering Theory (Gribov-Glauber Theory)~\cite{VNG}
is used. For nucleus-nucleus collisions, the Multiple Scattering Theory allows to consider this interactions as the
superposition of independent nucleon-nucleon interactions. 

At very high energies, the contribution of enhanced Reggeon diagrams (percolation effects) becomes important,
leading to a new effect,
the suppression of the inclusive density of secondaries~\cite{CKTr} into the central
(midrapidity) region. This effect corresponds to a significant fusion of the primarily produced quark-gluon strings.

The QGSM provides a succesful thorough description of multiparticle production processes in 
hadron-hadron~\cite{KaPi,Sh,ACKS,AMPS,MPS}, hadron-nucleus~\cite{KTMS,Sh1,AMSHpA}, 
and nucleus-nucleus~\cite{Sha,Shab,JDDS,AMSHpL} collisions, for a wide energy region. 
In partcular, the inclusive spectra of charged pions, kaons, nucleons, and $\Lambda$ hyperons, were correctly
described in the cited papers.  

The production of multistrange hyperons, $\Xi^-$ (dss), and $\Omega^-$ (sss), has special interest in 
high energy particle and nuclear physics.
Since the initial-state colliding projectiles contain no strange valence quarks,
all particles in the final state with non-zero strangeness quantum number should have been created
in the process of the collision. 
This makes multistrange baryons a valuable probe in
understanding the particle production mechanisms in high energy collisions.
A remarkable feature of strangeness production is that the production of each additional
strange quark featuring in the secondary baryons, i.e., the production rate of secondary 
$B(qqs)$ over secondary $B(qqq)$, then of $B(qss)$ over $B(qqs)$, and, finally,
of $B(sss)$ over $B(qss)$, is affected by one universal strangeness suppression factor, $\lambda_s$:
\begin{equation}
\lambda_s = \frac{B(qqs)}{B(qqq)} = \frac{B(qss)}{B(qqs)} = \frac{B(sss)}{B(qss)} \;,
\end{equation}
together with some simple quark combinatorics~\cite{AnSh,CS,AKMS}.

In the present paper, we compare the results of the standard QGSM calculations, with the experimental
data on yields of $p$, $\Lambda$, $\Xi$, and $\Omega$ baryons, and the corresponding antibaryons, in
nucleus-nucleus collisions, for a wide energy region, going from SPS up to LHC ranges. 

We also consider the ratios of multistrange to strange antihyperon
production in nucleus-nucleus collisions with different centralities, at CERN-SPS and RHIC energies.
In the standard formulation of QGSM, it is assumed that each secondary particle is produced by the
independent fragmentation of one single quark-gluon string.

Let us define: 
\begin{equation}
R(\overline{\Xi}^+/\overline{\Lambda}) = \frac{dn}{dy}(A+B\to\overline{\Xi}^++X)/ \frac{dn}{dy}
(A+B\to\overline{\Lambda}+X) \;,
\end{equation} 
\begin{equation}
R(\overline{\Omega}^+/\overline{\Lambda}) = \frac{dn}{dy}
(A+B\to\overline{\Omega}^++X)/ \frac{dn}{dy}
(A+B\to\overline{\Lambda}+X) \;.
\end{equation} 
The produced antihyperons, $\overline{\Xi}^+$ and $\overline{\Omega}^+$, contain  
valence antiquarks newly produced during the collision.
The ratios in eqs. (2) and (3) are reasonably described by QGSM when
a relatively small number of incident nucleons participate in the collision
(nucleon-nucleus collisions, or peripheral nucleus-nucleus collisions).

The number of quark-gluon strings (cut pomerons) in nucleus-nucleus
collisions increases with the centrality of the collision, but if the secondaries are independently 
produced in a single quark-gluon string, the ratio of yields of different 
particles should not depend on the centrality.

However, in this paper we show that in central nucleus-nucleus collisions already at 158 GeV/c per nucleon,
the experimental yield of multistrange hyperons is significantly larger then in peripheral collisions,
indicating that the ratios in eqs. (2) and (3) strongly depend on the centrality of the collision.
This effect decreases with the growth of the initial energy.
\vspace{-0.3cm}

\section{Other Models}

Already in ref.~\cite{Bocquet}, the UA1 Collaboration found, by measuring the strangeness yields in $p\overline{p}$
collisions at $\sqrt{s}$~=~630~GeV, that the strangeness suppression factor seemed to show a tendency to increase slowly with
$\sqrt{s}$, effect that it was conjectured it could be due to the incresase in the radiation of gluons which fragmnet into
$s\overline{s}$ pairs.

On the other hand,
by using the event generator LUCIAE to analyse the data by the NA35 Collaboration on the $\overline {p}$
and $\overline{\lambda}$ yields, in $pp$ and central sulphur-nucleus collisions at 200A GeV, Sa and Tai indicated~\cite{SaTai}
that those data might imply the reduction of strangeness suppression in ultrarelativistic nucleus-nucelus collisions
comparing to the nucleon-nucleon case at the same energy (see details on the event generator PACIAE, the updated version of
LUCIAE, in refs.~\cite{PACIAE,PACIAE_1}). Since the hadronic rescattering forcefully plays a role in the deficit of strangeness
suppression, the authors then claimed the necessity of going further in the study of the reduction of strangeness suppression,
before using it as a signal of quark-gluon plasma (QGP). 

The NeXus approach is used in ref.~\cite{Drescher} to analyse strangeness production in $pp$ collisions at SPS and RHIC energies,
to find that strangeness production is suppressed in proton-proton collisions at SPS energy, as compared to $e^+e^-$ annihilation,
while strangeness production in high energy $pp$ and $e^+e^-$ collisions agree quantitatively. The authors claim that the
strangeness suppression in $pp$ collisions at SPS is a consequence of the limited-space available in string fragmentation
at these energies. Thus, if the hadron mass is small as compared to the typical string energy, the hadron multiplicity ratios
(e.g. strange to not-strange), reach asymptotic values, and a further increase of the string energy leads only to an overall
increase of the produced hadron multiplicity, leaving their relative ratio unchanged. If, on the contrary, the string mass becomes
comparable to the hadron masses (when the energy of the collision is smaller), the production of the heavy hadrons is suppressed
due to the very limited phase space available.

In~\cite{Tounsi}, Tounsi and Redlich claimed that the pattern of enhancement of strange and multistrange baryons observed by the
WA97 Collaboration could be understood on the basis of the significant role of canonical suppression in the enhancement of the
strangeness production.
By presenting quantitative predictions for the relative enhancement of $\Lambda$, $\Xi$, and $\Omega$ yields in the energy range
from $\sqrt{s}$~=~8.7~GeV to $\sqrt{s}$~=~130~GeV,
they showed that in the canonical approach the enhancement of the strangeness production is a decreasing
function of collision energy, where the enhancement of $\Omega$, $\Xi$, and $\lambda$ would be of the order 100, 20, and 3,
respectively. Strangeness enhancement is already seen at low energies, and found to be a decreasing function of collision energy
in a set of data on the $K^+/\pi^+$ ratio, in AA relative to $pp$ collisions. Such a behavior with the energy of the collision could
also be expected for multistrange baryons, and, in fact, a statistical model implementing canonical strangeness conservation
explains~\cite {Tounsi1} the WA97 pattern and predicts that the enhancement is a decreasing function of collision energy.
As an example, Tounsi and Redlich predict that at RHIC enhancement of $\Omega$ is smaller than at SPS, in contrast with
the predictions by the UrQMD model of an enhancement at RHIC larger by a factor of four than at SPS~\cite{UrQMD}.
It was also stressed the fact that the NA57 Collaboration data~\cite{WA97} which measure the yield per participant in Pb+Pb
collisions relative to p+Pb and p+Be collisions show a saturation in the enhancement of the strangeness production for a number
of participant nucleons
$N_{part}>$100, while results of the NA57 Collaboration~\cite{NA5722} indicated an abrupt change of $\overline{\Xi}^+$ enhancement
for lower values of $N_{part}$. Similar behavior had been previously seen on the $K^+$ yield measured by the NA22 Collaboration
in Pb+Pb collisions~\cite{NA5222}.

In ref.~\cite{Becattini}, where it is pointed out that the correlation between the enhancement in the strangeness production
and the $J/\Psi$ anomalous
suppression patters observed in heavy ion collisions at top SPS energy $\sqrt{s}$~=~17.2~GeV,
if studied as a function of the transverse size of the interaction region, makes considerably stronger the
case for SPS being right at the onset of QGP formation, it is also found that the strangeness suppression factor varies
considerably between the most central and the most peripheral bin in Pb+Pb collisions, being significantly smaller in central
C+C and Si+Si than in Pb+Pb collisions, confirming the results of a previous analysis in ref.~\cite{Cleymans}. Thus, the strangeness
suppression factor is larger in central light-ion collisions than in the most peripheral heavy-ion collisions, where it approaches
the value found in $pp$ collisions. These analyses raised the question on whether a regular behavior exists as a function of the
system size in the strangeness production pattern in heavy ion collisions, either by taking $\lambda_s$ as a function of the
fraction of struck participants in the Glauber Model~\cite{Cleymans}, either by looking for an underlying physical mechanism
implying the undersaturation of the strange phase space, with $\lambda_s $ becoming an effective parametrization of a more
complicated physical picture (i.e. see~\cite{Becattini1}).

In ref.~\cite{Fochler}, two different types of transport models, the microscopic transport model UrQMD, and a stochastic transport
model provided by an elaborated transport cascade, were used to show that both dynamically account
for the suppression in the yields of rare strange particles in $\pi\pi$ collisions, provided
in the canonical formulation of the Boltzmann equation, and associated with the exact conservation of an U(1)-charge.
The reason is that any kaon in the system requires the existence of a corresponding
anti-kaon due to strangeness conservation, thus the probability of finding a particle anti-particle pair turns into a highly
correlated conditional probability, the enhanced annihilitaion probability then leading to the canonical suppression in the kaon
yields, behavior automatically included in the considered transport models.

As a matter of fact, and as noticed in~\cite{Lobanov}, the value of strangeness suppression factor $\lambda_s$ is measured and
calculated by different methods. Thus, it has been considered as constant in energy range by Malhotra and Orava~\cite{Orava},
while there are other analyses which accept the decrease of strangeness suppression with the increase of the collision
energy~\cite{Bocquet,Wroblewski}, and new Monte-Carlo event generators have been developped where $\lambda_s$ is energy
dependent~\cite{Long}.

The PYTHIA~6.4 predictions for strange particle spectra were calculated by Lobanov and collaborators~\cite{Lobanov} with new ATLAS
tunes, to show that the analysis of particle yields in $pp$ collisions at the LHC energies demonstrates the smaller strangeness
suppression comparing with experiments at lower energies. Thus, the experimental data by the ALICE~\cite{ALICE22}, CMS~\cite{CMS22},
and ATLAS~\cite{ATLAS22} collaborations show a large increase in the measured production cross section of strange particles as
$\sqrt{s}$ increases from 0.9 to 7~TeV.

The UA1 Collaboration found a value of $\lambda_s$=0.29 from their measurements on the ratio $K/\pi$ at energy $\sqrt{s}$~=~630~GeV,
in agreement with previous measurements~\cite{Orava}, and with UA5 and CDF experiments. Then they made energy dependence fitting
for $\lambda_s$, and found a value $\lambda_s$=0.31 at the LHC energy of 14~TeV. From LHC data~\cite{ALICE22,CMS22,ATLAS22},
in the range $\sqrt{s}$~=~0.9 to 7~TeV, in the central rapidity region, Lobanov and collaborators obtain an estimation
of $\lambda_s$=0.39. They conclude that LHC data show an increase of strangeness production (or smaller strangeness suppression),
with respect to previous experimental data in lower energy ranges.

Moreover, H.~Satz signaled~\cite{Satz} that in a hydrodynamical scenario the (universal) strangeness suppression factor
for hadroproducion in high energy $pp$, $pA$, and $AA$ collisions increases from 0.5 to 1.0 in a narrow temperature range
around the quark-hadron transition temperature, and strangeness suppression disappears with the onset of color deconfinement,
and with the onset of the full equilibrium resonance gas behavior.

At this point, one has to note that the strangeness suppression factor, $\lambda_s$, accounts for the suppression,
in the same process, of the production of strange particles with respect to the production of non-strange particles, while the term
strangeness enhancement indicates the, in principle expected enhancement, when going from less to more energetic/central collisions,
of strange particle production. Finally, if one observes that this experimental enhancement is less than the theoretically predicted,
one talks of suppression, or decrease, of the strangeness enhancement.

Regarding approaches specifically based on the QGSM formalism, different Monte Carlo event generators built from the QGSM framework
have been used to quantitatively describe the main properties of multiparticle events in hadron-nucleus and nucleus-nucleus collisions at
high energies. One of the first theoretically consistent translations of the QGSM to a Monte Carlo event generator is due
to N.S.~Amelin and L.V.~Bravina~\cite{AmelinBravina,AmelinBravina1}.

In further developments, S.G.~Mashnik and collaborators presented the Los Alamos
version of the QGSM, realized in the high energy code LAQGSM~\cite{LAQGSM}, able to describe both particle and nucleus induced reactions
at energies up to 1~TeV/nucleon. LAQGSM differs from QGSM by replacing the preequilibrium and eveporation parts with the new physics
from CEM2k~\cite{CEM2k}, and it has a number of improvements in the cascade and Fermi break-up models. LAQGSM is particularly suited
to be used is cosmic-ray applications~\cite{Mashnik}.

In ref.~\cite{QGSJETII}, S.~Ostapchenko introduced the construction of the Monte Carlo generator QGSJET-II, for high energy hadronic and nuclear
collisions, in which interactions are treated in the framework of the Reggeon Field Theory (RFT), enhanced Pomeron diagrams are taken into
account, and where both soft and semihard contributions to the parton dynamics are accounted for. Since it is based on the RFT formalism,
QGSJET-II shares most of its assumptions, like the validity of the Abramovski-Gribov-Kancheli (AGK) rules and eikonal vertices for
Pomeron-hadron vertices. It also neglects energy-momentum correlations between multiple scattering processes at the amplitude level.

In ref.~\cite{Bleibel}, the Monte Carlo realization of the QGSM in ref.~\cite{AmelinBravina,AmelinBravina1} was used to study the main
characteristics of $pp$ interactions at energies from $\sqrt{s}$~=~200~GeV to LHC energy $\sqrt{s}$~=~14~TeV, and it is shown that QGSM
favors violation of Feynman scaling in the central rapidity region and its preservation in the fragmentation regions, it holds extended
longitudinal scaling at LHC, and it presents further violation of the Koba-Nielsen-Olesen (KNO) scaling in multiplicity distributions.

In particular concerning the description of strangeness production by using a QGSM-based Monte Carlo, in ref.~\cite{Amelin_PRC47}
the Monte Carlo QGSM~\cite{AmelinBravina,AmelinBravina1} was used to study the strangeness production in proton and heavy ion collisions
at 200A GeV, the shape of rapidity and transverse mass distributions being well reproduced, both in peripheral and central heavy ion collisions.
In fact, for the hadron-nucleus case, the Monte Carlo QGSM successfully describes not only the shape of distributions, but the absolute yield
of neutral strange particles, as well. The description gets less reliable for strange particle abundances produced in heavy colliding systems,
what it is seen in the model results for central S+S collsions. The authors note that in QGSM, secondary interactions at energies lower than
the threshold for strange particle production lead even to decreasing ratios strange/negative, while experimental ratios have the opposite trend:
Strangeness abundance grows with increasing negative multiplicity (decreasing impact parameter), the most drastic difference being observed for
$\overline{\Lambda}$ production. In this regard, the NA35 Collaboration data~\cite{NA35,NA35_1} are in an unique position, since a strangeness
enhancement with decreasing impact parameter had not been observed  at lower AGS-BNL energies, or for the same 200A GeV energy, with lighter projectile.
To give a qualitative estimate of the physical consequences of including collective string-string interactions to improve the description of central
heavy-ion collisions, also in this paper some preliminary results of the Monte Carlos String Fusion Model (SFM)~\cite{ABP,ABPZP} are presented.
\vspace{-0.3cm}

\section{Production of Secondaries in the QGSM}

The QGSM~\cite{KTM,KaPi} provides
quantitative predictions of different features of multiparticle production,
in particular, of the inclusive densities of different secondaries, both in
the central, and in the beam fragmentation regions. In QGSM, high energy
hadron-nucleon collisions are implemented through the exchange
of one or several Pomerons, all elastic and inelastic processes resulting
from cutting between (elastic) or through (inelastic) Pomerons~\cite{AGK}.

Each Pomeron corresponds to a cylindrical diagram, and thus, when cutting
one Pomeron, two showers of secondaries are produced (see figs.~1a and~1b).
\begin{figure}[htb]
\vskip -3.3cm
\hskip -1.2cm
\includegraphics[width=.6\hsize]{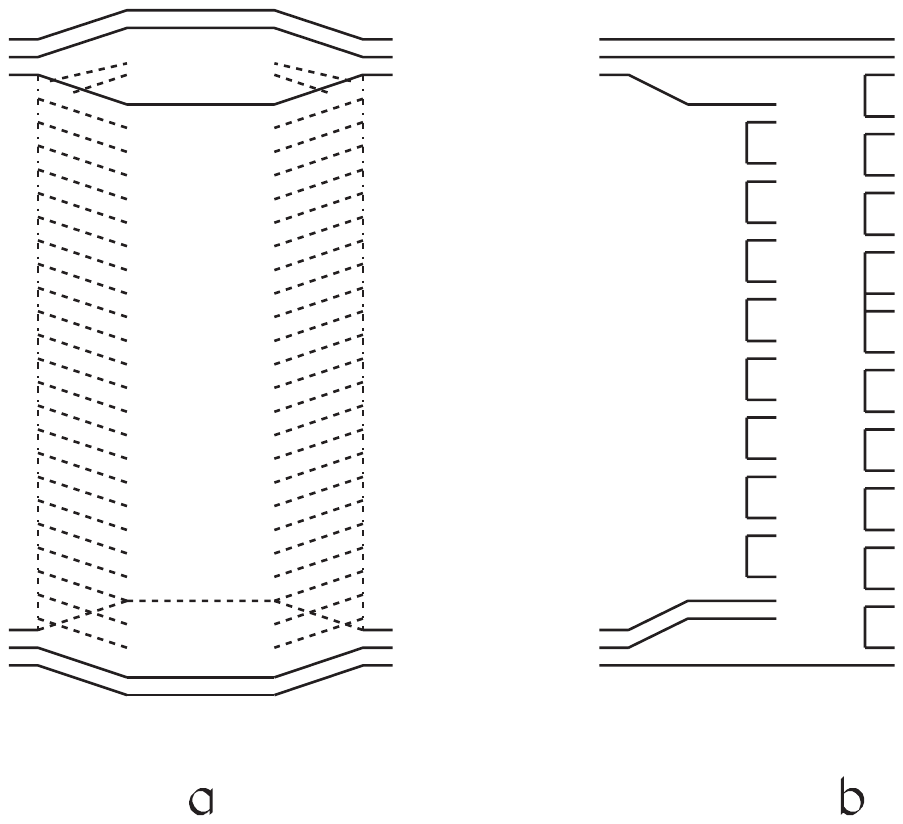}
\hskip -1.2cm
\includegraphics[width=.6\hsize]{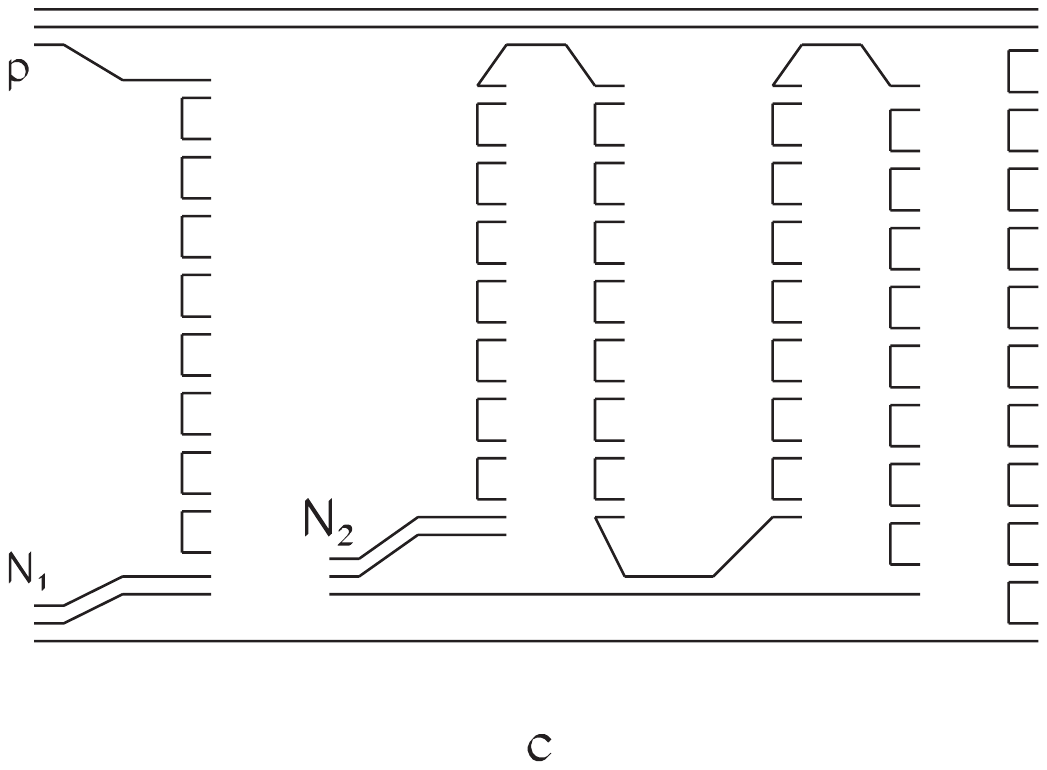}
\vskip -3.4cm
\caption{\footnotesize
(a) Cylindrical diagram representing the Pomeron exchange within the DTU
classification. Quarks are shown by solid lines; (b) Cut of the cylindrical
diagram corresponding to the single-Pomeron exchange contribution in inelastic
$pp$ scattering; (c) Diagram corresponding to the inelastic interaction of an
incident proton with two target nucleons, $N_1$ and $N_2$, in a $pA$
collision.}
\end{figure}

It was assumed in the QGSM that the Pomeron corresponding to the cylinder-type diagrams is a simple pole with
$\alpha_P(0)>1$ (supercritical Pomeron), with a trajectory given by
\begin{eqnarray}
\alpha_P(t)=1+\Delta+\alpha'_P\cdot t\ , \Delta > 0\ ,
\end{eqnarray}
the values used in QGSM for the intercept, $\Delta$, and for the slope, $\alpha'_P$,
are~\cite{Sh}:
\begin{eqnarray}
\Delta=0.139 , \alpha'_P=0.21\ .
\end{eqnarray}
These values have been the ones used to describe the experimental data, both at SPS, and at LHC energies,
and, for the sake of consistency, they have not been changed in the different calculations performed by QGSM for
more many years, though their fine tunning could better the agreement of the QGSM results with some of the LHC
data by a small percentage.

For such a supercritical Pomeron, higher terms of the topological expansion associated with exchange by several
Pomerons in the $t$ channel are also important. This is due to the fact that, though the exchange by $n$ Pomerons is
$\sim 1/(N^2)^{n}$~\cite{Sh}, it is enhanced by the factor $(s)^{n\Delta}$~\cite{TerMartirosyan_73}. Thus,
it is necessary to sum many terms of the topological expansion at very high energy. The $s$-channel discontinuities
of these diagrams are related to processes of production of $2k$ ($k\le n$) chains of particles.

The AGK-cutting rules~\cite{AGK} make it possible to determine the cross sections for $2k$-chain (string) production
(with any number of uncut Pomerons) if the contributions of all $n$-Pomeron exchanges to the forward elastic
amplitude are known. These contributions can be calculated using the Reggeon diagram technique~\cite{Gribov_1967}.
In principle, in most calculations diagrams of the eikonal type are taken into account.

In this model, the cross sections of $2k$-chain production $\sigma_k$ have the form~\cite{TerMartirosyan_73}
\begin{eqnarray}
\sigma_k(\xi)=\frac{\sigma_P}{k_z}\cdot\left[1-exp(-z)\sum^{k-1}_{i=0}\frac{z^i}{i!}\right] \;,\ k\le1\ ,
\end{eqnarray}
where
\begin{eqnarray}
\sigma_P=8\pi\gamma_P\ exp(\Delta\xi),\ z=\frac{2C\gamma_P}{R^2+\alpha^{\prime}\xi}\cdot exp(\Delta\xi),
\ \xi=ln(s/s_0)\;  
\end{eqnarray}
(see more details in ref.~\cite{K20}).

The total interaction cross section has the form
\begin{eqnarray}
\sigma^{(tot)}&=&\sum^{\infty}_{k=0}\sigma_k(\xi)=\sigma_P\cdot f\left(\frac{z}{2}\right)\ \;,\nonumber\\
f\left(\frac{z}{2}\right)&=&\sum^{\infty}_{n=1}\frac{(-z)^{n-1}}{n\cdot n!}\ \; .
\end{eqnarray}

For superhigh energies, when $\xi >> 1$, $\sigma^{(tot)}(\xi)$ has a Froissart-type behavior
\begin{eqnarray}
\sigma^{(tot)}(\xi)\sim\frac{8\pi\alpha^{\prime}_P\Delta}{C}\xi^2\ \; .
\end{eqnarray}
This type of behavior for the total cross section is common to a wide class of models with $\alpha_P>1$.
Inclusive cross sections and multiplicity distributions can be obtained in this approach by summing over hadronic
production for all processes of $2k$-chain formation.

The inclusive spectrum of a secondary hadron $h$ is then determined by the convolution
of the diquark, valence quark, and sea quark distribution functions, $u(x,n)$ (every one
normalized to unity), in the incident particles, with the fragmentation functions $G^h(z)$
for quarks and diquarks to contribute to the production of the secondary hadron $h$.
Both the distribution functions and the fragmentation functions are constructed by using the
well established Reggeon counting rules~\cite{Kai}.

In particular, when $n > 1$, i.e. in the case of multipomeron
exchange, the distributions of valence quarks and diquarks are softened due
to the appearence of a new contribution from sea quark-antiquark pairs. 

The details of the model are presented in~\cite{KTM,KaPi,Sh,ACKS}.
The average number of exchanged Pomerons slowly increases with the energy of the collision.

For a nucleon target, the inclusive rapidity, $y$, or Feynman-$x$, $x_F$,
spectrum of a secondary hadron $h$ has the form~\cite{KTM}:
\begin{equation}
\frac{dn}{dy}\ = \
\frac{x_F}{\sigma_{inel}}\cdot \frac{d\sigma}{dx_F}\ = \
\sum_{n=1}^\infty w_n\cdot\phi_n^h (x) + w_D \cdot\phi_D^h (x) \ ,
\end{equation}
where the functions $\phi_{n}^{h}(x)$ determine the contribution of diagrams
with $n$ cut Pomerons, $w_n$ is the relative weight of this diagrams, and the
term $w_D \cdot\phi_D^h (x)$ accounts for the contribution of diffraction
dissociation processes.

For $pp$ collisions:
\begin{equation}
\phi_n^{h}(x) = f_{qq}^{h}(x_{+},n) \cdot f_{q}^{h}(x_{-},n) +
f_{q}^{h}(x_{+},n) \cdot f_{qq}^{h}(x_{-},n) +
2(n-1)f_{s}^{h}(x_{+},n) \cdot f_{s}^{h}(x_{-},n)\ \  ,
\end{equation}
\begin{equation}
x_{\pm} = \frac{1}{2}[\sqrt{4m_{T}^{2}/s+x^{2}}\pm{x}]\ \ ,
\end{equation}
where $m_T = \sqrt{m^2 + p^2_T}$ is the transverse mass of the produced hadrons, and 
$f_{qq}$, $f_{q}$, and $f_{s}$, 
correspond to the contributions of diquarks, valence, and sea quarks (antiquarks), respectively,
The contributions of the incident particle and the target proton depend on the variables $x_{+}$ 
and $x_{-}$, respectively~\cite{KTM,KTM82,K20,Kai}.
 
The first two terms in Eq.~(5) correspond to chains with valence quarks (diquarks) at the ends, 
while the last term is connected to chains stretched between sea quark-antiquark.
The expression of the functions $f$ in Eq.~(5) is determined by 
the convolution of the diquark and quark distribution functions, with the fragmentation functions, e.g.,
\begin{equation}
f_i^h(x_{\pm},k)\ =\ \int\limits_{x_{\pm}}^1u_i(x_1,k)G_i^h(x_{\pm}/x_1) dx_1\ \;,
\end{equation}
where $i$= $qq$ diquarks, $q$ valence quarks, and $\bar{q}$ valence antiquarks, and $q_s$ sea quarks
($\bar{q_s}$ sea antiquarks),
respectively.

In the calculation of the inclusive spectra of secondaries produced in $pA$ collisions, we should consider
the possibility of one or several Pomeron cuts in each of the $\nu$ blobs of the proton-nucleon inelastic
interactions (see Fig.~1c).

It is also essential to take into account all diagrams with every possible Pomeron configuration, and its
corresponding permutations~\cite{KTM, Kaid, Shab}. The diquark and quark distribution functions and the
fragmentation functions are the same as in the case of a proton-nucleon, $pN$, interaction.

The process shown in Fig.~1c satisfies~\cite{BT,Weis,Sh3,Jar} the condition
that the absorptive parts of the hadron-nucleus amplitude are determined by
the combination of the absorptive parts of the hadron-nucleon amplitudes.

For nucleus-nucleus collisions, in the fragmentation region of
the projectile we use the approach presented in refs.~\cite{Sha,Shab,JDDS}, where the beam
of independent nucleons of the projectile interacts with the target nucleus,
what corresponds to the rigid target approximation~\cite{Alk} of the
Glauber Theory. Correspondingly, in the target fragmentation region the
beam of independent target nucleons interacts with the projectile nucleus.
The results obtained in this approach for the two fragmentation regions coincide in the
central region. When the initial energy is not very high, the corrections due to the implementation
of energy conservation play here a very important role. This approach was used in~\cite{JDDS}
to obtain a successful description of the yields of $\pi^{\pm}$, $K^{\pm}$, $p$, and $\overline{p}$,
in Pb+Pb collisions at 158 GeV per nucleon.

The superposition picture of QGSM gives a reasonable description~\cite{MPS,KTMS,JDDS} of the 
inclusive spectra on nuclear targets at energies $\sqrt{s}$~=~14$-$30~GeV.
At RHIC energies the situation drastically changes: while the spectra of
secondaries produced in $pp$ collisions are described by QGSM rather well~\cite{MPS},
the RHIC experimental data for $Au$+$Au$ collisions~\cite{Phob,Phen} give clear evidence
of suppression effects that reduce the midrapidity inclusive density by about a factor two,
when compared to the predictions based on the superposition picture~\cite{Shab,CMT,AP}.
This reduction can be explained by the inelastic screening
corrections connected to multipomeron interactions~\cite{CKTr} (see Fig.~2).
\begin{figure}[htb]
\vskip -3.8cm
\hskip 2.0cm
\includegraphics[width=.7\hsize]{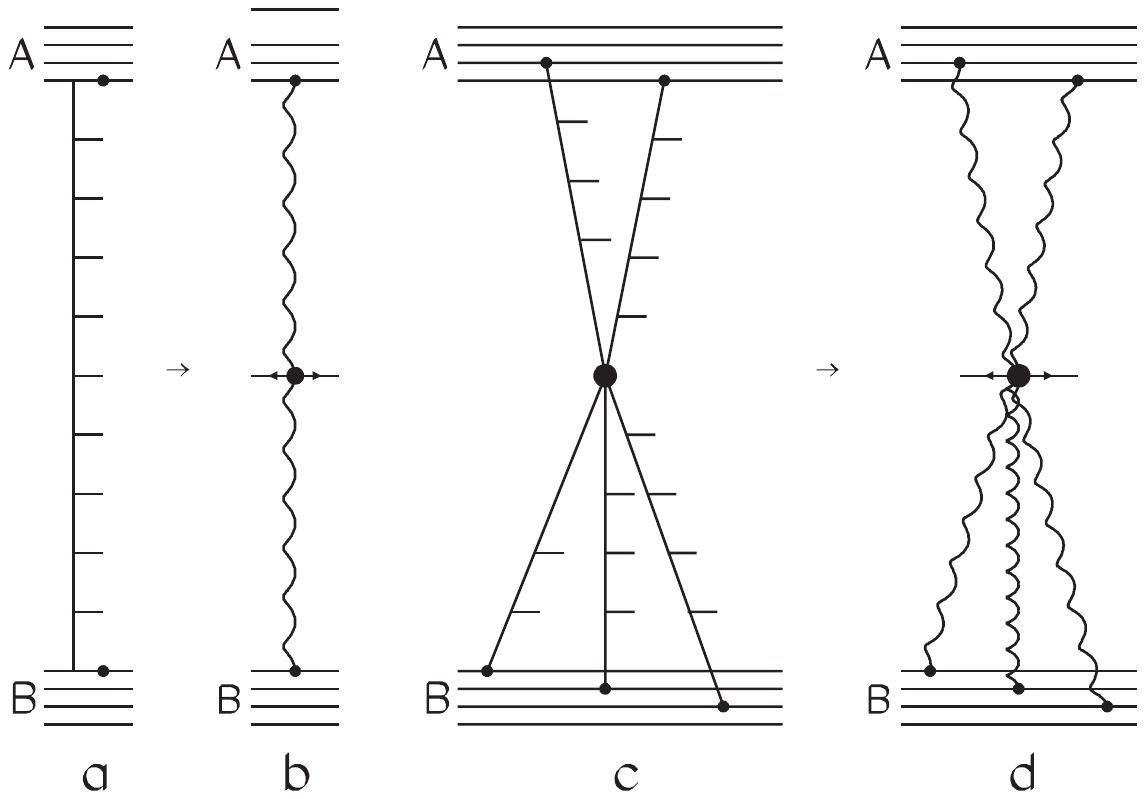}
\vskip -3.8cm
\caption{\footnotesize
(a) Multiperipheral ladder diagram; (b) Inclusive cross section corresponding to diagram in (a);
(c) Diagram with fusion of several ladders; (d) Inclusive cross section corresponding to diagram in (c).}
\end{figure}
All these diagrams are proportional to the squared longitudinal form factors of both colliding nuclei~\cite{CKTr}. 

Though all numerical estimates are model dependent, since the numerical weight of the contribution of the 
multipomeron diagrams remains rather unclear due to the many unknown vertices in these diagrams, in some models
the number of unknown parameters can be reduced. As an example of this, in reference~\cite{CKTr}
the Schwimmer model~\cite{Schw} was used to obtain consistent numerical predictions.

In order to account for the screening effects in the QGSM, it is technically more 
simple~\cite{MPS} to consider the maximal number of Pomerons, $n_{max}$, emitted by one nucleon in the central
region. Then, after they are cut, these Pomerons lead to the different final states. Finally, the contributions
of all diagrams with $n \leq n_{max}$ are accounted for, as at lower energies.
The unitarity constraint also affects the emission of larger numbers
of Pomerons, $n > n_{max}$, but due to fusion in the final state (at the quark-gluon
string stage), the cut of $n > n_{max}$ Pomerons results in the same final
state as the cut of $n_{max}$ Pomerons.

Actually, the largest number of Pomerons one accounts for in QGSM depends on (grows with) the energy, and it is fixed
in the inelastic cross section as $10^{-3}/$(all sum), that is, $10^{-3}$ times the sum of the Pomerons in all considered
diagrams, for the calculation at one given energy.

By taking this approximation, all QGSM calculations become rather simple, straightforward, 
and very similar to those in the percolation approach~\cite{ABP,ABPZP,ABFP}. 
\vspace{-0.3cm}

\section{Comparison of QGSM calculations with experimental data}

\subsection{Fixed Target Energy Data}

The experimental data on proton and hyperon productiom were reasonably described in~\cite{AMSHpA}.
The data for p+Be and p+Pb collisions were also described in the same reference, but it appears
that the value of the strange suppression parameter $\lambda_s$=0.25 would have to be  slightly
increased to get a better comparison with the experimental data. In the present paper we use the value
$\lambda_s$=0.32, that provides good results for the description of the experimental data in
references~\cite{AMSHnphi,AMSHKst}. 
The results of our calculation for p+Be and p+Pb collisions are presented in Table~1, where it is
apparent the consistent agreement with the experimental data.

In the case of Pb+Pb collisions at 158 GeV/c per nucleon, we compare the experimental data
on $dn/dy$ for central collisions ($\mid y\mid\leq 0.5$), measured by
NA49~\cite{NA49pbp,NA49c,NA49d,NA49b,NA49a,BM}, NA57~\cite{NA57b,NA57}, 
and WA97~\cite{WA97} collaborations, with the corresponding QGSM predictions. 
The comparison for inclusive densities $dn/dy$ of secondary proton and hyperons are shown in Table·~2, where
one can see that the experimental data obtained by each of the collaborations are, in general, compatible with
those by the others.

\begin{center}
\vskip -3pt
\begin{tabular}{|c|c|c|} \hline
Process & dn/dy, $|y|\leq 0.5$ (Experimental Data) & dn/dy (QGSM) \\
\hline

 p + Be $\to \Lambda$ & $0.0334 \pm 0.0005 \pm 0.003$ &0.0357  \\

 p + Be $\to \overline{\Lambda}$  & $0.011 \pm 0.0002 \pm 0.001$ & 0.0128  \\  \hline

 p + Pb $\to \Lambda$  & $0.060 \pm 0.002 \pm 0.006 $ & 0.0678  \\

 p + Pb $\to \overline{\Lambda}$ & $0.015 \pm 0.001 \pm 0.002$  &0.0173 \\ \hline

 p + Be $\to \Xi^-$  & $0.0015 \pm 0.0001 \pm 0.0002$  & 0.00261  \\

 p + Be $\to \overline{\Xi}^+$  & $0.0007 \pm 0.0001 \pm 0.0002$ &0.00113 \\ \hline

 p + Pb $\to \Xi^-$ & $0.0030 \pm 0.0002 \pm 0.0003$ &0.00336   \\

 p + Pb $\to \overline{\Xi}^+$  & $0.0012 \pm 0.0001 \pm 0.0001$&0.00113  \\  \hline 

 p + Be $ \to \Omega^-$  & $0.00012 \pm 0.00006 \pm 0.00002$  &  0.000143  \\

 p + Be $\to \overline{\Omega}^+$  & $0.00004 \pm 0.00002 \pm 0.00001$  &0.0000453  \\ \hline

 p + Pb $ \to \Omega^-$ & $0.00022 \pm 0.00008 \pm 0.00003$  &0.000184  \\

 p + Pb $\to \overline{\Omega}^+$ & $0.00005 \pm 0.00003 \pm 0.00002$   &0.000058  \\ \hline

\hline
\end{tabular}
\end{center}
Table 1: {\footnotesize Experimental data by the NA57 Collaboration~\cite{NA57} on inclusive
densities $dn/dy$ of hyperon production in p+Be and p+Pb central $\mid y\mid\leq 0.5$ collisions
at proton-nucleon 158 GeV/c, together with the results of the QGSM calculations ($\lambda_s$ = 0.32}).

The values in Table~2 state the reasonable description of the experimental data for secondary $p$, $\Lambda$,
and $\overline{\Lambda}$, obtained in QGSM, by using the value $\lambda_s$=0.32, for the strangeness suppession
factor, $\lambda_s$.

On the contrary, for the case of multistrange hyperon production of ${\Xi}^-$, $\overline{\Xi}^+$, $\Omega^-$,
and $\overline{\Omega}^+$, the values in Table~2 show that in the frame of the QGSM a correct description of the
experimental data can only be obtained by using significant larger values of $\lambda_s$, i.e. by implementing
a weaker suppression of strangeness production. This fact is a clear indication of a significant violation of the
quark combinatorial rules~\cite{AKMS,CS}.

Actually, the quark combinatorial rules assume that every strange quark can be picked up with the same probability,
i.e. with the same value of $\lambda_s$, but comparison in Table~2 
demonstrates the opposite picture: the value of the parameter $\lambda_s$ for the cases of 
${\Xi}^-$, $\overline{\Xi}^+$, $\Omega^-$, and $\overline{\Omega}^+$ production appears to be more than 
3 times larger than for the cases of $\Lambda$ and $\overline{\Lambda}$ production.

In Fig.~3 we show the comparison of the QGSM prediction with the experimental data~\cite{NA57}
on the dependence of the ratios $\overline{\Omega}^+$/$ \overline{\Lambda}$ (left panel) and  
$\overline{\Xi}^+$ to $\overline{\Lambda}$ (right panel), on the number of wounded nucleons, $N_{\omega}$,
in Pb+Pb collisions at 158~GeV/c per nucleon.
The number of wounded nucleons, $N_{\omega}$, is directly related to the centrality of collisions. Thus,
small values of $N_w$ correspond to peripheral collisions (large impact parameter), while large values
of $N_w$ corresponds to central collisions (small impact parameter).

\begin{center}
\vskip -5pt
\begin{tabular}{|c|c|c|c|c} \hline
Process & Centrality & dn/dy, $|y|\leq 0.4$ (Experimental Data) & dn/dy (QGSM) \\ \hline 

Pb+Pb  $\to p$ &0$-$5\% &$29.6  \pm 0.9~\cite{NA49pbp}$ & 30.29 \\ \hline

Pb+Pb $\to \Lambda$ & 0$-$10\%&$9.5 \pm 0.1 \pm 1.0~\cite{NA49b,BM}$  &6.64 \\ \hline

Pb+Pb $\to \Lambda$ & 0$-$10\%&$10.9 \pm 1.0 \pm 1.3~\cite{NA49c,BM}$& \\ \hline

Pb+Pb $\to \Lambda$ & 0$-$10\% & $13.7 \pm 0.9$~\cite{BM,WA97} & \\ \hline

Pb+Pb $\to \overline{\Lambda}$& 0$-$10\%  & $1.24 \pm 0.03 \pm 0.13~\cite{NA49b,BM}$ & 1.65 \\ \hline

Pb+Pb $\to \overline{\Lambda}$& 0$-$10\%  & $1.62 \pm 0.16 \pm 0.2~\cite{NA49c,BM}$  & \\ \hline

Pb+Pb $\to \overline{\Lambda}$ & 0$-$10\% & $1.8 \pm 0.2$~\cite{BM,WA97} & \\ \hline

Pb+Pb $\to \Xi^-$& 0$-$5\% & $2.08 \pm 0.09 \pm 0.21~\cite{BM,NA57}$ & 1.458 \\ \hline

Pb+Pb $\to \overline{\Xi}^+$& 0$-$5\% & $0.51 \pm 0.04 \pm 0.05~\cite{BM,NA57}$ & 0.482 \\ \hline

Pb+Pb $\to \Xi^-$ & 0$-$10\% & $1.5 \pm 0.01$~\cite{BM,WA97} & 1.34 \\ \hline

Pb+Pb $\to \overline{\Xi}^+$ & 0$-$10\% & $0.37 \pm 0.06$~\cite{BM,WA97}& 0.443 \\ \hline

Pb+Pb $\to \Xi^-$ & 0$-$10\% & $1.43 \pm 0.33 \pm 0.16$~\cite{NA49a,BM} & 1.34 \\ \hline

Pb+Pb $\to \Xi^-$ & 0$-$10\% & $1.44 \pm 0.10 \pm 0.15$~\cite{NA49b,BM}& 1.34 \\ \hline

Pb+Pb $\to \overline{\Xi}^+$ & 0$-$10\% & $0.31 \pm 0.03 \pm 0.03$~\cite{NA49b,BM} & 0.443 \\ \hline

Pb+Pb $\to \Omega^-$& 0$-$5\% & $0.31 \pm 0.07 \pm 0.05~\cite{BM,NA57}$  & 0.246 \\ \hline

Pb+Pb $\to \overline{\Omega}^+$& 0$-$5\%  & $0.16 \pm 0.04 \pm 0.02~\cite{BM,NA57}$ & 0.111 \\ \hline

Pb+Pb $\to \Omega^-$& 0$-$10\% & $0.14 \pm 0.03 \pm 0.01~\cite{NA49d,BM}$  & 0.197 \\ \hline

Pb+Pb $\to \overline{\Omega}^+$& 0$-$10\%  & $0.07 \pm 0.03 \pm 0.01~\cite{NA49d,BM}$  & 0.071 \\ \hline

Pb+Pb $\to \Omega^-$& 0$-$11\% & $0.259 \pm 0.037 \pm 0.026~\cite{BM,NA57b}$& 0.197 \\ \hline

Pb+Pb $\to \overline{\Omega}^+$& 0$-$11\%  & $0.129 \pm 0.0122 \pm 0.013~\cite{BM,NA57b}$ &  0.071 \\ \hline

\end{tabular}
\end{center}
Table 2: {\footnotesize The comparison of QGSM predictions with the experimental data on
midrapidity yields of $p$, $\overline{p}$, strange $\Lambda$, $\overline{\Lambda}$,
and multistrange ${\Xi}^-$, $\overline{\Xi}^+$, $\Omega^-$, and $\overline{\Omega}^+$,
produced at different centrality percentage in Pb+Pb collisions at
energy 158 GeV/c per nucleon,
measured by the NA49~\cite{NA49pbp,NA49c,NA49d,NA49b,NA49a,BM}, NA57~\cite{NA57b,NA57}, 
and WA97~\cite{WA97} collaborations. Experimental data presented in the review~\cite{BM}
are also considered. The values of the strangeness suppression factor, $\lambda_s$,
used in the different QGSM calculations presented in this table, are the following:
$\lambda_s$=0.32 (for $p$ and $\Lambda$
production), $\lambda_s$=0.7 (for $\Xi$ production), and $\Lambda_s$=0.75 (for $\Omega$ production).}
\vskip 0.75cm

\begin{figure}[htb]
\vskip -3.5cm
\hskip -1.5cm
\includegraphics[width=.65\hsize]{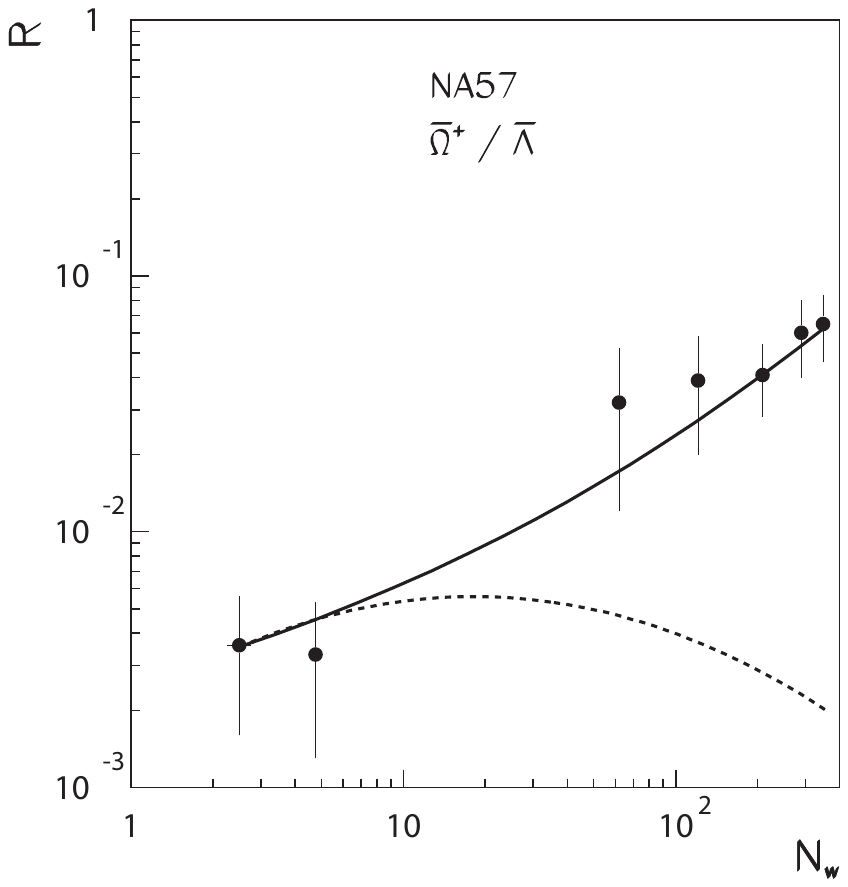}
\hskip -2.75cm
\includegraphics[width=.65\hsize]{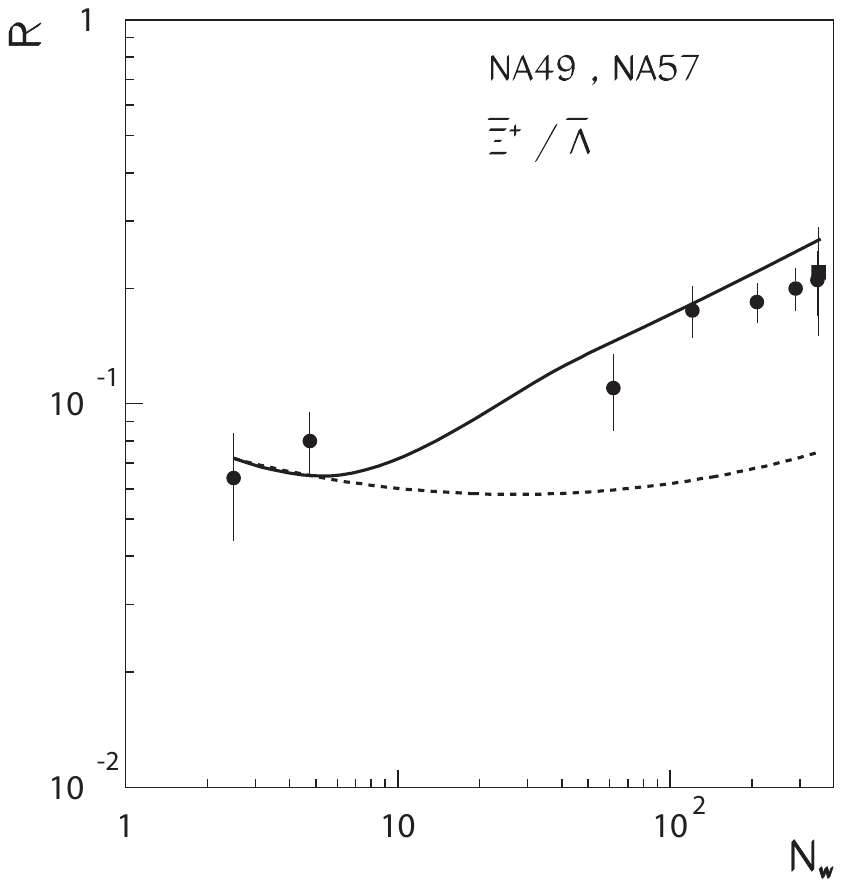}
\vskip - 3.5cm
\caption{\footnotesize
Ratios of $\overline{\Omega}^+$/$\overline{\Lambda}$ (left panel), and
of $\overline{\Xi}^+$/$\overline{\Lambda}$ (right panel) as functions of the
number of wounded nucleons, $N_w$. The experimental data of Pb+Pb collisions
for different numbers of wounded nucleons, $N_w$ (different centralities)
measured by the NA57 Collaboration (points), and by the NA49 Collaboration (squares)
are presented, and compared with the corresponding QGSM predictions.}
\end{figure}

The necessity of using different effective values of the strangeness suppression factor $\lambda_s$, to correctly
describe the production, both of strange, and of multistrange hadrons at SPS, directly appears when looking at the
corresponding experimental data. In particular, this need is made apparent by the increase with centrality of the
experimental ratios of the yields $\overline{\Omega}^+/\overline{\Lambda}$, and $\overline{\Xi}^+/\overline{\Lambda}$,
measured in Pb+Pb collisions (see Fig.~3), since all other parameters in the model remain invariant.

At small values of $N_w$ the ratio is practically equal to that in the cases of p+Be and p+Pb
collisions, and all they can be correctly described by the QGSM by using a value the of the strangeness suppression
parameter, $\lambda_s$=0.32. Then, the experimental ratio increases rather fast
with the incresing value of $N_w$, i.e. when we move from peripheral to central Pb+Pb collisions.
This behavior is reasonably reproduced by the full line in Fig.~3, that it has been calculated with the value
$\lambda_s$=0.32 at its left end, and with the values of $\lambda_s$ shown in Table~2 at its right end.
 
The results of QGSM calculations with a constant value $\lambda_s$=0.32, disregarding of the value of the number of
wounded nucleons (centrality), $N_{\omega}$, are shown in Fig~3 by dashed lines.

Obviously, for small values of $N_w$, both curves coincide, but when $N_w$ increases the full line also increases 
in agreement with the data, while the dashed line is practically constant, with exception of small corrections 
mainly connected to energy conservation. This dashed line shows a very significant disagreement with the experimenatl data
at large values of $N_w$.

The difference between the full and the dashed lines for the ratio $\overline{\Omega}^+$/$ \overline{\Lambda}$
in a very central event is of about one order of magnitude.
Such a large difference comes from the fact that the cross section for $\overline{\Omega}$ production 
is proportional to ${\lambda_s}^3$.
Meanwhile, since the cross section for $\overline{\Xi}^+$ production
is proportional to ${\lambda_s}^2$, the difference between full and dashed curves for the ratio
$\overline{\Xi}^+$/$ \overline{\Lambda}$ is not so large.

This behavior in which the value of the strangeness suppression factor $\lambda_s$ increases with the
value of $N_w$ (centrality), indicates that the simple quark combinatorial rules~\cite{AKMS,CS} are not valid
for central collisions of heavy nuclei.

In figs.~4 and 5, the rapidity dependence of $p$, $\overline{p}$, $ \Lambda$, $\overline{\Lambda}$ (Fig. 4),
and of ${\Xi}^-$, $\overline{\Xi}^+$, ${\Omega}^-$, $\overline{\Omega}^+$ (Fig. 5) baryons production
in central Pb+Pb collisions with initial momentum 158 GeV/c per nucleon, are shown, and compared with the QGSM predictions
(full and dashed lines). The full and dashed lines have been calculated for baryon and antibaryon production by using
the corresponding values of the strangeness suppression parameter, ${\lambda_s}$, given in Table~2.

Both in figs.~4 and~5, one can appreciate a reasonable agreement of the QGSM predictions with the experimental data for
the production of all secondary bayons.

\begin{figure}[htb]
\vskip -3.2cm
\hskip -1.5cm
\includegraphics[width=.65\hsize]{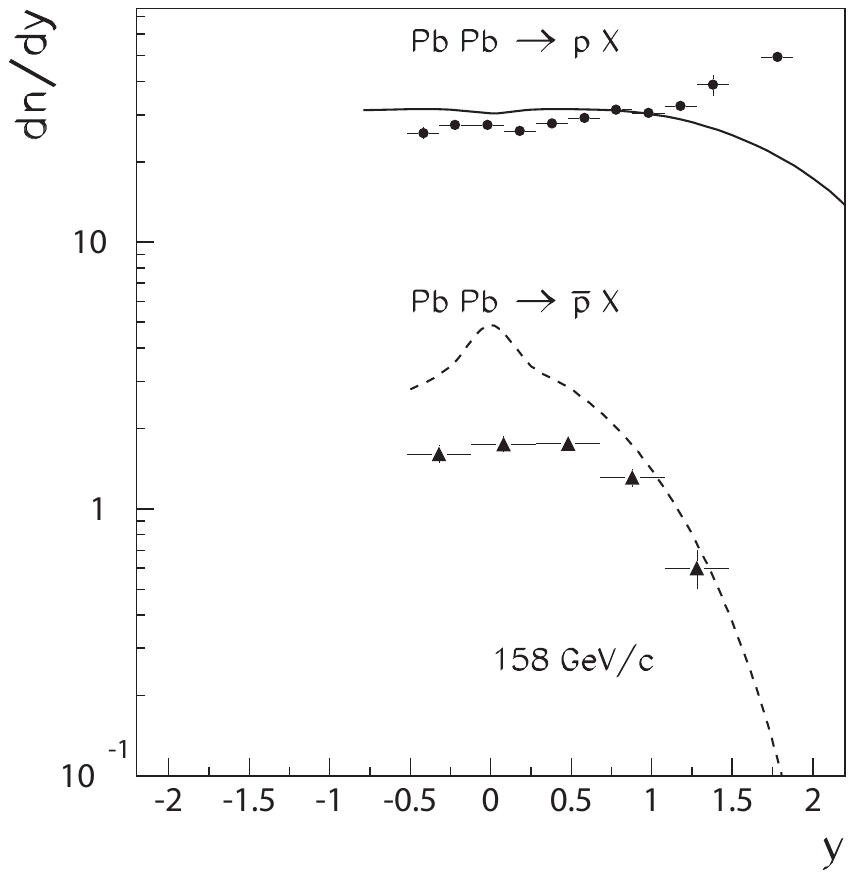}
\hskip -2.75cm
\includegraphics[width=.65\hsize]{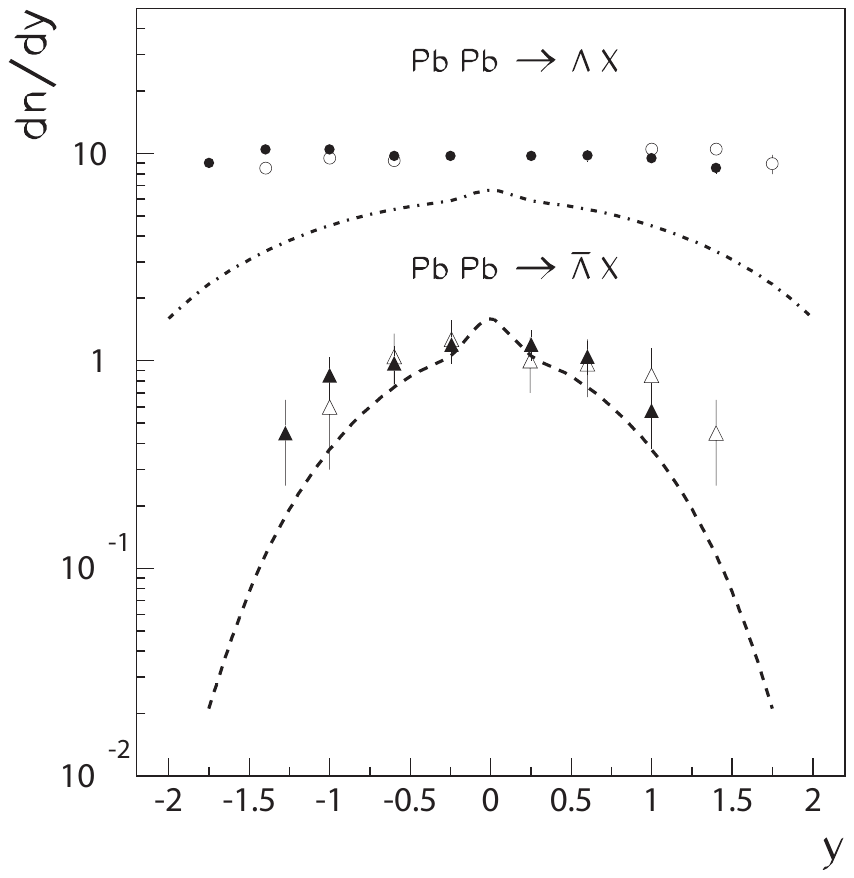}
\vskip - 3.5cm
\caption{\footnotesize
Rapidity dependence of $dn/dy$ for $p$, $\bar p$ (left panel), and for $\Lambda$,
$\overline{\Lambda}$ (right panel) productions. 
The experimental data of Pb+Pb collisions measured by the NA49 Collaboration at 158 GeV/c per nucleon~\cite{NA49b}
are presented, and compared with the QGSM predictions (solid and dashed lnes).}
\vskip -3.0cm
\hskip -1.5cm
\includegraphics[width=.65\hsize]{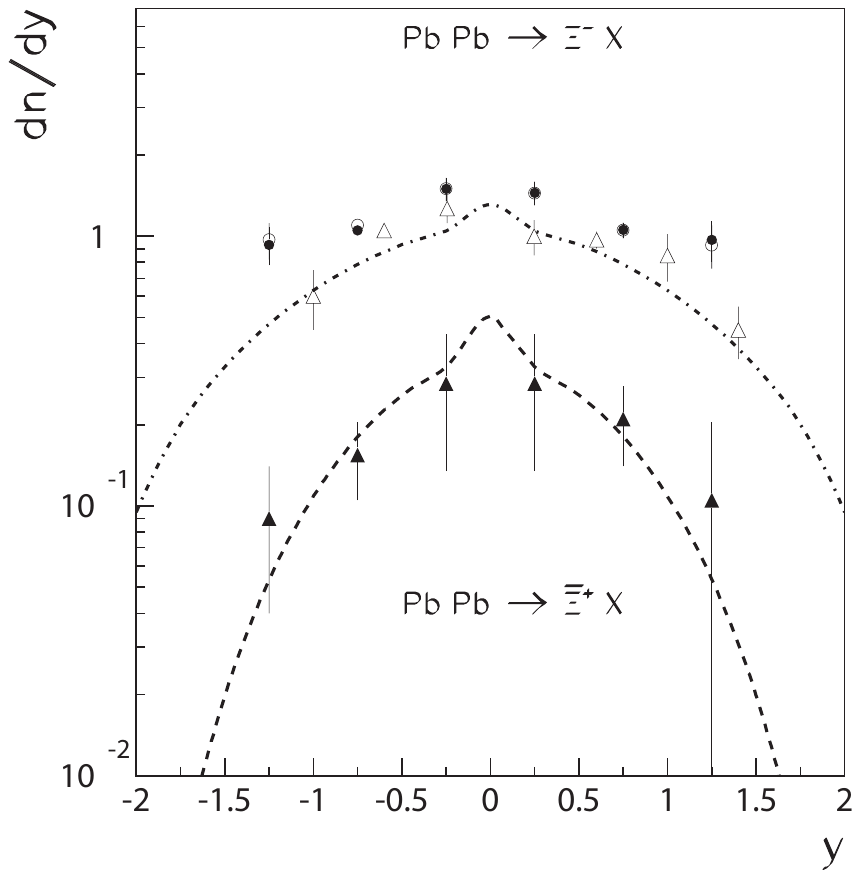}
\hskip -2.75cm
\includegraphics[width=.65\hsize]{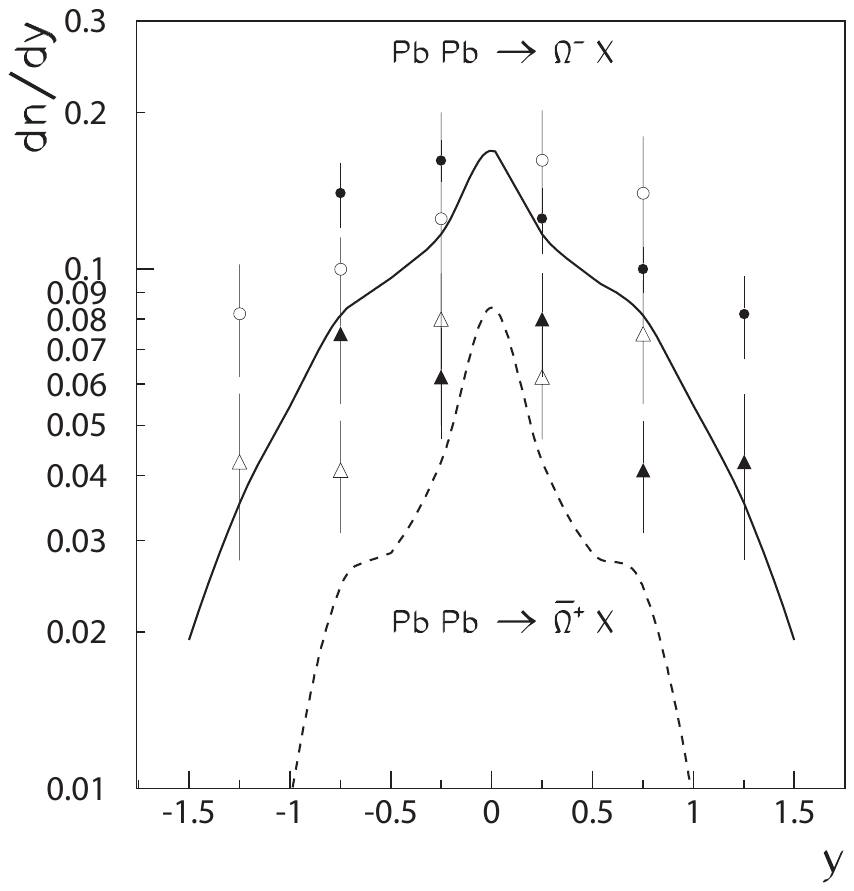}
\vskip -3.5cm
\caption{\footnotesize
Rapidity dependence of $dn/dy$ for $\Xi^-$, $\overline{\Xi}^+$ (left panel), and $\Omega ^-$, $\overline{\Omega ^+}$ 
(right panel) productions. The experimental data of Pb+Pb collisions measured by the NA49 Collaboration at 158 GeV/c per
nucleon~\cite{NA49d} are presented, and compared to the QGSM predictions (solid and dashed lines).}
\vskip -0.5cm
\end{figure}

\newpage

\subsection{STAR Collaboration Data}

Now we consider the experimental data on midrapidity densities of protons and hyperons  
in Au+Au and Cu+Cu collisions measured by the STAR~\cite{STAR3,STAR2,STARp,STAR,STAR1}
and PHENIX~\cite{PHENIX,PHENIXp} collaborations at RHIC energies, and we compare them with the results of the
corresponding QGSM calculations. The comparison is presented in Table~3, were one can appreciate that, in general,
the agreement of the QGSM predictions with the experimental data in this energy region is quite reasonable.

Again, we see here that the value of $\lambda_s$ for multistrange hyperon production is larger than for the case
of $\Lambda$ and $\overline{\Lambda}$ production. Now this difference is not so large as it is for collisions at
lower energies (see Table~2). Moreover, it seems that the difference in the value of the parameter $\lambda_s$
decreases with the growing of the energy, meaning that the violation of the quark combinatorial rules becomes less
important for high energy collisions.

In Fig.~6 we present the comparison of the QGSM predictions with experimental data on the $N_{\omega}$ dependence
of the ratios $\overline{\Omega}^+$/$\overline{\Lambda}$ (left panel), and $\overline{\Xi}^+$/$\overline{\Lambda}$
(right panel), measured by the STAR Collaboration in the midrapidity region, at $\sqrt{s}$~=~62.4~GeV.
Similarly as
in Fig.~3, the left end of the full line here was calculated with a value $\lambda_s$=0.32,
while for the right end the values of $\lambda_s$ presented in Table~3 were used.
The dashed line
was calculated with a constant value of $\lambda_s$=0.32. 

The same calculations for the ratio of $\overline{\Xi}^+$/$\overline{\Lambda}$ in Cu+Cu collisions
at $\sqrt{s}$~=~200~GeV, are shown in Fig.~7. Here we see again that the violation of the quark combinatorial rules 
decreses with the growth of the energy of the collision,
the difference between the full and dashed lines being of the order of the experimental error bars,
at $\sqrt{s}$~=~200~GeV.

\footnotesize{
\begin{center}
\begin{tabular}{|c|c|c|c|c|c|c|}\hline
Process & $\sqrt{s}$ (GeV) & Centrality & dn/dy (Experimental Data) & dn/dy (QGSM) & $\lambda_s$ \\ \hline

Au+Au $\to p $ & 62.4 &0$-$5\% & $29.0 \pm 3.8 ~\cite{STARp} $ &22.826 & 0.32 \\ \hline

Au+Au $\to \overline{p}$ & 62.4 &0$-$5\% & $13.6 \pm 1.7~\cite{STARp} $ & 15.115 & \\ \hline

Au+Au $\to \Lambda$ & 62.4 &0$-$5\% & $15.7 \pm 0.3 \pm 2.3~\cite{STAR} $ &9.221 & 0.32 \\ \hline

Au+Au $\to \overline{\Lambda}$ & 62.4 &0$-$5\% & $8.3 \pm 0.2 \pm 1.1~\cite{STAR} $ & 6.258 & \\ \hline

Au+Au $\to \Xi^-$ &62.4 &0$-$5\% & $ 1.63 \pm 0.09 \pm 0.18 ~\cite{STAR}$&1.406 & 0.4 \\ \hline

Au+Au $\to \overline{\Xi}^+$ &62.4 &0$-$5\% & $ 1.03 \pm 0.00 \pm 0.11~\cite{STAR}$ & 0.996 & \\ \hline

Au+Au $\to \Omega^-$ &62.4 & 0$-$20\% & $0,212 \pm 0.028 \pm 0.018~\cite{STAR}$ &0.222 & 0.55 \\ \hline

Au+Au $\to \overline{\Omega}^+$ &62.4 & 0$-$20\% & $0,167 \pm 0.027 \pm 0.015~\cite{STAR}$ & 0.154 & \\ \hline

Au+Au $\to p $ & 130. &0$-$5\% & $28.2 \pm 3.1 ~\cite{STARp}$ &27.450 & 0.32 \\ \hline

Au+Au $\to p $ & 130. &0$-$5\% & $28.7 \pm 0.9 ~\cite{PHENIXp}$ &27.450 &  \\ \hline

Au+Au $\to \overline{p}$ &130. &0$-$5\% & $20.0 \pm 2.2~\cite{STARp} $ & 22.723 & \\ \hline

Au+Au $\to \overline{p}$ &130. &0$-$5\% & $20.1 \pm 1.0~\cite{PHENIXp} $ & 22.723 & \\ \hline

Au+Au $\to \Lambda$ & 130. & MB & $4.8 \pm 0.3 ~\cite{PHENIX} $ &3.379 & 0.32 \\ \hline 

Au+Au $\to \overline{\Lambda}$ & 130. & MB & $4.3 \pm 0.7~\cite{PHENIX}$ &2.861 & \\ \hline 

Au+Au $\to \Lambda$ & 130. &0$-$5\% & $17.3 \pm 1.8 \pm 2.8~\cite{PHENIX}$ &13.070 & 0.32 \\ \hline 

Au+Au $\to \Lambda$ & 130. &0$-$5\% & $17.0 \pm 0.4 \pm 1.7~\cite{STAR} $ &  & \\ \hline

Au+Au $\to \overline{\Lambda}$ & 130. &0$-$5\% & $12.7 \pm 1.82 \pm 2.0~\cite{PHENIX}$ & 11.053 & \\ \hline

Au+Au $\to \overline{\Lambda}$ & 130. &0$-$5\% & $12.3 \pm 0.3 \pm 1.2S~\cite{STAR}$ & & \\ \hline

Au+Au $\to \Xi^-$ & 130. & 0$-$10\% & $2.0 \pm 0.14 \pm 0.2$~\cite{STAR3} & 1.976 & 0.39 \\ \hline

Au+Au $\to \overline{\Xi}^+$ & 130 & 0$-$10\% & $1.70 \pm 0.12 \pm 0.17$~\cite{STAR3} & 1.754 & \\ \hline

Au+Au $\to \Omega^-$+ $\overline{\Omega}^+$ & 130. & 0$-$10\% & $0.55 \pm 0.11 \pm 0.06$~\cite{STAR3} &0.544 & 0.48 \\ 
\hline

Cu+Cu $\to \Lambda$ &200 &0$-$10\% & $4.68 \pm 0.45~\cite{STAR1} $ & 4.200 & 0.32 \\ \hline

Cu+Cu $\to \overline{\Lambda}$ &200 &0$-$10\% & $3.79 \pm 0.37~\cite{STAR1} $ & 3.772 & \\ \hline

Cu+Cu $\to \Xi^-$ &200 & 0$-$10\% & $0.62 \pm 0.08~\cite{STAR1} $ &0.559 & 0.33 \\ \hline

Cu+Cu $\to \overline{\Xi}^+$ & 200 & 0$-$10\% & $0.52 \pm 0.08$~\cite{STAR1} & 0.507 & \\ \hline

Cu+Cu $\to  \Omega^-$+ $\overline{\Omega}^+$ & 200. & 0$-$10\% & $0.141 \pm 0.017$~\cite{STAR1} & 0.140 & 0.39 \\ 
\hline

Au+Au $\to p $ & 200. &0$-$5\% & $34.7 \pm 4.4 ~\cite{STARp}$ & 31.450 & 0.32 \\ \hline

Au+Au $\to \overline{p}$ &200. &0$-$5\% & $26.7 \pm 3.4~\cite{STARp} $ & 27.765 & \\ \hline

Au+Au $\to \Lambda$ & 200 & 0$-$5\% & $14.8 \pm 1.5$~\cite{STAR1} & 15.762 & 0.32 \\ \hline

Au+Au $\to \Lambda$ &200 &0$-$5\% & $16.7 \pm 0.2 \pm 1.1~\cite{STAR2} $ &  & \\ \hline

Au+Au $\to \overline{\Lambda}$ &200 &0$-$5\% & $11.7 \pm 0.9~\cite{STAR1} $ & 14.152 & \\ \hline

Au+Au $\to \overline{\Lambda}$ &200 &0$-$5\% & $12.7 \pm 0.2 \pm 0.9~\cite{STAR2} $  & &  \\ \hline

Au+Au $\to \Xi^-$  &200 &0$-$5\% & $ 2.17 \pm 0.06 \pm 0.19 ~\cite{STAR2}$& 2.173 & 0.34 \\ \hline

Au+Au $\to \overline{\Xi}^+$ &200 &0$-$5\% & $ 1.83 \pm 0.05 \pm 0.20~\cite{STAR2}$ & 1.962 &  \\ \hline

Au+Au $\to \Omega^-$ + $\overline{\Omega}^+$ &200. &0$-$5\% & $0.53 \pm 0.04 \pm 0.03~\cite{STAR2}$ & 0.525 & 0.4 \\ 
\hline
\end{tabular}
\end{center}
}}
\vskip -0.15cm

\noindent
Table 3: {\footnotesize Experimental data on dn/dy by the STAR Collaboration~\cite{STAR1} on $p$, $\bar p$, $\Lambda$,
$\overline{\Lambda}$, $\Xi^-$, $\overline{\Xi}^+$ and $\Omega^-$, $\overline{\Omega}^+$ production
in central Cu+Cu and Au+Au collisions at RHIC energies, together with the corresponding QGSM results.}

\newpage

\newpage
\begin{figure}[htb]
\vskip -2.0cm
\hskip -1.5cm
\includegraphics[width=.65\hsize]{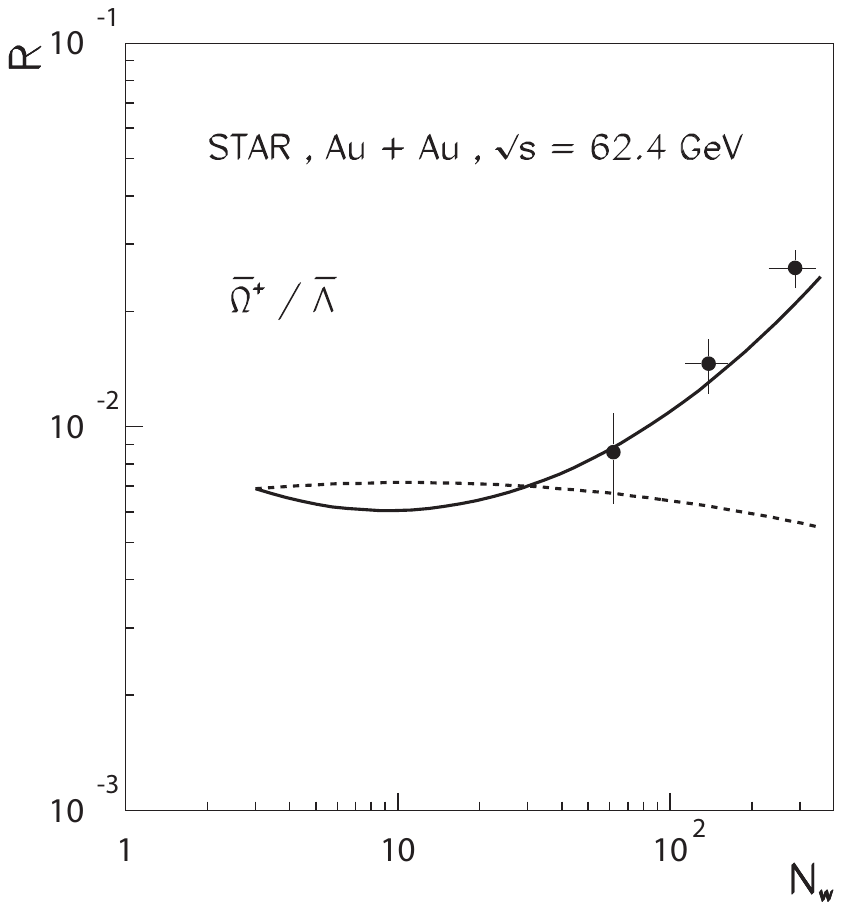}
\hskip -2.75cm
\includegraphics[width=.65\hsize]{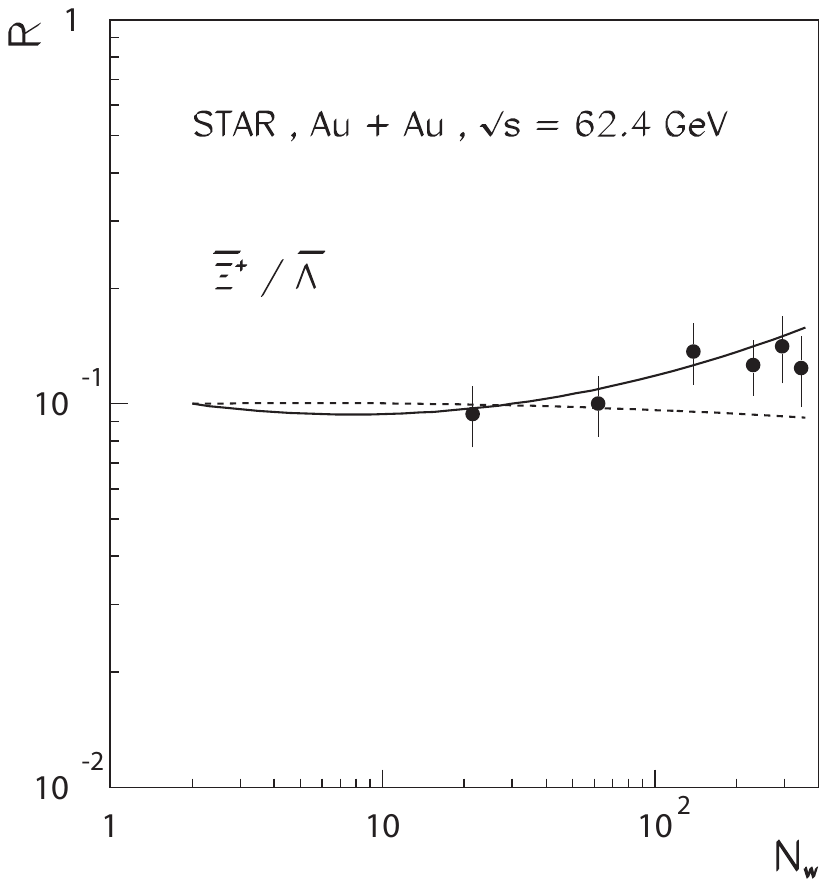}
\vskip -3.25cm
\caption{\footnotesize The experimental points obtained by the STAR Collaboration on the ratios of
$\overline{\Omega}^+$/$\overline{\Lambda}$ (left panel), and $\overline{\Xi}^+$ to $\overline{\Lambda}$ (right panel),
in Au+Au collisions at $\sqrt{s}$~=~62.4~GeV, at different centralities,
as a function of the number of wounded nucleons, $N_w$,
together with the results of the corresponding QGSM calculations (solid curves).}
\vskip -3.0cm
\hskip 2.0cm
\includegraphics[width=.7\hsize]{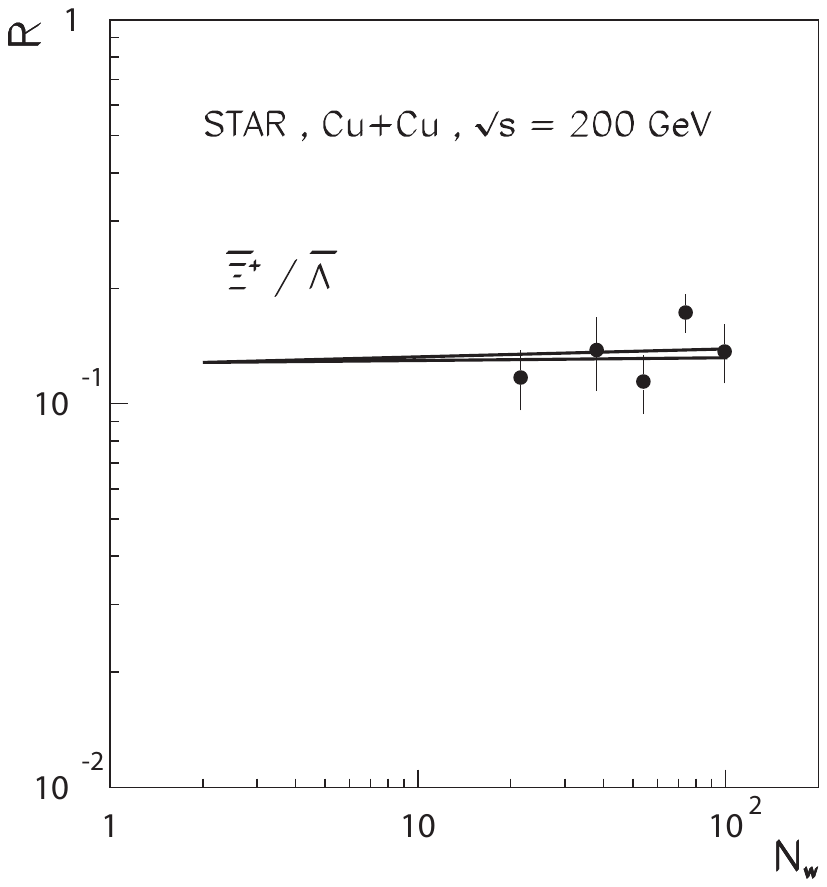}
\vskip -3.75cm
\caption{\footnotesize
The experimental points obtained by the STAR Collaboration on the ratios $\overline{\Xi}^+$/$\overline{\Lambda}$
in Cu+Cu collisions at $\sqrt{s}$~=~200~GeV, at different centralities,
as a function of the number of wounded nucleons, $N_w$,
together with the results of the corresponding QGSM calculations.}
\vskip -3.0cm
\end{figure}
\vskip -0.5cm
\newpage

\noindent
\subsection{LHC experimental data}

In Table~4 we consider the experimental data on $p$, $\overline{p}$, and ${\Xi}^-$, $\overline{\Xi}^+$, ${\Omega}^-$, 
$\overline{\Omega}^+$ production in central Pb+Pb collisions at $\sqrt{s}$~=~2.76~Tev, and of $p+\overline{p}$ production
in central Pb+Pb collisions at $\sqrt{s}$~=~5.02~Tev,
measured by the ALICE Collaboration~\cite{ALICEPbp276,ALICEms276,ALICEPbp5}, at the CERN LHC.
{\small{
\begin{center}
\vskip -0.5cm
\begin{tabular}{|c|c|c|c|c|c|c|} \hline
Process & $\sqrt{s}$ (TeV) & Centrality & dn/dy (Experimental Data) & dn/dy (QGSM) & $\lambda_s$ \\ \hline

Pb+Pb  $\to p$ & 2.76 &0$-$5\% &$34  \pm 3~\cite{ALICEPbp276}$ & 34.604 & 0,32 \\ \hline

Pb+Pb  $\to \overline{p}$ & 2.76 &0$-$5\% &$ 33 \pm 3~\cite{ALICEPbp276}$ & 33.898 & \\ \hline

Pb+Pb  $\to p + \overline{p}$ & 5.02 &0$-$5\% & $74.56 \pm 0.06 \pm 3.75~\cite{ALICEPbp5}$ & 77.71 & 0,32 \\ \hline

Pb+Pb  $\to \Xi^-$ & 2.76 &0$-$10\% &$3.34 \pm 0.06 \pm 0.24~\cite{ALICEms276}$ & 3.357 & 0,32 \\ \hline

Pb+Pb  $\to \overline{\Xi}^+$ & 2.76 &0$-$10\% & $3.28 \pm 0.06 \pm 0.23~\cite{ALICEms276}$ & 3.317 &  \\ \hline

Pb+Pb  $\to \Omega^-$ &2.76  & 0$-$10\% & $0,58 \pm 0.04 \pm 0.09$~\cite{ALICEms276} & 0.606& 0.38 \\ \hline

Pb+Pb  $\to \overline{\Omega}^+$ &2.76  & 0$-$10\% & $0,60 \pm 0.05 \pm 0.09$~\cite{ALICEms276}& 0,601 & \\ \hline

\end{tabular}
\end{center}
Table 4: {\footnotesize Experimental data on dn/dy by the ALICE Collaboration of $p$, $\bar p$ central
production at $\sqrt{s}$~=~2.76~TeV~\cite{ALICEPbp276}, of $p$+$\bar p$ central production
at $\sqrt{s}$~=~5.02~TeV~\cite{ALICEPbp5}, ${\Xi}^-$, and of $\Xi^-$, $\overline{\Xi}^+$, ${\Omega}^-$,
and $\overline{\Omega}^+$ production in central Pb+Pb collisions at $\sqrt{s}$~=~2.76~TeV per nucleon~\cite{ALICEms276},
and the comparison with the results of the corresponding QGSM calculations.}
}}
\vskip 0.15cm

\noindent
Here we can see that the strangeness suppression parameter $\lambda_s$ for ${\Xi}^-$ and $\overline{\Xi}^+$
production becomes smaller than at RHIC energies, taking the standard value $\lambda_s$=0.32.
In the case of ${\Omega}^-$ and $\overline{\Omega}^+$ production, the value of $\lambda_s$ also decreases
with respect to the RHIC energy range.
Thus we see that the unusually large values of $\lambda_s$ for central Pb+Pb collisions at 158~GeV/c per nucleon,
monotonically decrease with the increase of the initial energy of the collision.

Actually, this is the point we want to stress: though it seems to contradict the accepted believe
that the QGP is easier to get produced in nuclear collisions at RHIC or LHC energies, the experimental data on strange
and multistrange particle production show that the value of the strangeness suppression factor to be taken in QGSM to
correctly describe those experimental data, it is lower at RHIC and LHC energy ranges, than at SPS (i.e. a smaller
enhancement of strangeness production at RHIC and LHC, when compared with that at SPS). It has to be inderlined that
this surprising behavior it is directly shown by the experimental data, and it is not the result, or feature, of any
theoretical description of the experimental data.
\vspace{-0.3cm}

\newpage

\section{Conclusion}

We consider the production of hyperons in collisions on nuclear targets, in the frame of the QGSM formalism, and
we find that one unexpected scenario appears when comparing the QGSM predictions with the available experimental data.
Thus, while the experimental data on multistrange hyperon and antihyperon production in $p$-nucleus and in peripheral
nucleus-nucleus collisions can be reasonably described by the QGSM, by using for all baryons the same standard value of
the strange suppression parameter $\lambda_s$=0.32 (see Table~1 and figs.~3,~6, and~7), to get a correct description in
QGSM of multistrange hyperon and antihyperon production in central Pb+Pb collisions in the midrapidity region, a larger
value of $\lambda_s$ is needed (as it can be seen in Table~2). In particular, in Fig.~3 one can see that the experimental
probability of multistrange hyperon production in Pb+Pb collisions increases monotonically with the number of wounded
nucleons, $N_w$, i.e. when in nucleus-nucleus going from peripheral to very central collisions.

A similar situation also occurs in the case of Au+Au collisions at RHIC energies (see Table~3 and Fig.~6), while
in the case of Cu+Cu collisions this effect is practically negligible (see Table~3 and Fig.~7),
probably because the Cu nucleus is not heavy enough.

Actually, the experimental data on the production of multistrange hyperons in central collisions of heavy nuclei show
a very significant violation, essentially large at CERN-SPS energies~$\sqrt{s}$~=~17.3~GeV, and that it decreases with the
growth of the initial energy of the collision, of simple quark combinatorial rules~\cite{AnSh,CS,AKMS}, while, at the same
time, the corresponding data in proton-nucleus and in peripheral nucleus-nucleus collisions are in agreement with the
theoretical description based on those quark combinatorial rules.

We point out the remarkable experimental fact of the different strangeness suppression in multistrange hyperon production,
from central collisions of heavy nuclei to proton-nucleus and peripheral nucleus-nucleus collisions, and the decreasing of
this effect with the increase of the initial energy of the collision.

The full theoretical explanation, as well as the description of the corresponding microscopic mechanism, of this experimental
effect are still to be formulated, what it will be crucial to determine if, and under which conditions, the reduction
of strangeness suppression can be considered as an undisputed signature of the realization of the phase transition to QGP in
high energy nuclear collisions. 
\vskip 0.4cm

\noindent
{\bf Acknowledgements}

This paper was supported by Ministerio de Econom{\'i}a, Industria y Competitividad, Spain (Mar{\'i}a de Maeztu Unit
of Excellence MDM-2016-0692), and by Xunta de Galicia, Galiza-Spain (Consolidaci\'on e Estructuraci\'on 2021 GEC GI-2033-TEOFPACC).


\vspace{-0.3cm}

\newpage

\begin{thebibliography}{**}

\bibitem{KTM82} A.B. Kaidalov and K.A. Ter-Martorisyan, Phys. Lett. {\bf B117}, 247 (1982),
doi:10.1016/0370-2693(82)90556-1.

\bibitem{KTM} A.B. Kaidalov and K.A. Ter-Martirosyan, Sov. J. Nucl. Phys. {\bf 39}, 979 (1984),
Yad. Fiz. {\bf 39}, 1545 (1984); Sov. J. Nucl. Phys. {\bf 40}, 135 (1984), Yad. Fiz. {\bf 40}, 211 (1984).

\bibitem{Kaid} A.B. Kaidalov, Sov. J. Nucl. Phys. {\bf 45}, 902 (1987), Yad. Fiz. {\bf 45}, 1452 (1987).

\bibitem{K20} A.B. Kaidalov, Phys. Atom. Nucl. {\bf 66}, 1994 (2003), doi:10.1134/1.1625743,\\
Yad. Fiz. {\bf 66}, 2044 (2003), Proceedings of Modern Trends in Classical Approach: \\International Seminar Devoted
to 80th Birthday of K.A. Ter-Martirosyan, Moscow (Russian Federation), September 30th-October 1st (2002).

\bibitem{RV} G.C.~Rossi and G.~Veneziano, Nucl. Phys. {\bf B123}, 507 (1977), doi:10.1016/0550-3213(77)90178-X.

\bibitem{GMV} M.~Ciafalloni, G.~Marchesini and G.~Veneziano, Nucl. Phys. {\bf B98}, 472 (1975),
doi:10.1016/0550-3213(75)90502-7.

\bibitem{VNG} V.N. Gribov, Sov. Phys. JETP {\bf 29}, 483 (1969), Zh. Eksp. Teor. Fiz. {\bf 56}, 892 (1969).

\bibitem{CKTr} A. Capella, A. Kaidalov, and J. Tran Thanh Van, Acta Phys. Hung. {\bf A9}, 169 (1999),
e-Print:9903244[hep-ph].

\bibitem{KaPi} A.B.~Kaidalov and O.I.~Piskounova, Sov. J. Nucl. Phys. {\bf 41}, 816 (1985),
Yad. Fiz. {\bf 41}, 1278 (1985); Z. Phys. {\bf C30}, 145 (1986), doi:10.1007/BF01560688.

\bibitem{Sh} Yu.M. Shabelski, Sov. J. Nucl. Phys. {\bf 44}, 117 (1986), Yad. Fiz. {\bf 44}, 186 (1986).

\bibitem{ACKS} G.H. Arakelyan, A. Capella, A.B.~Kaidalov, and Yu.M.~Shabelski,
Eur. Phys. J. {\bf C26}, 81 (2002), doi:10.1007/s10052-002-0977-z,
e-Print:0103337[hep-ph]

\bibitem{AMPS} G.H. Arakelyan, C. Merino, C. Pajares, and Yu.M.~Shabelski,
Eur. Phys. J. {\bf C54}, 577 (2008), doi:10.1140/epjc/s10052-008-0554-1, e-Print:0709.3174[hep-ph].

\bibitem{MPS} C. Merino, C. Pajares and Yu.M.~Shabelski, Eur. Phys. J. {\bf C71}, 1652 (2011),
doi:10.1140/epjc/s10052-011-1652-z.

\bibitem{KTMS} A.B. Kaidalov, K.A. Ter-Martirosyan, and Yu.M.~Shabelski,
Sov. J. Nucl. Phys. {\bf 43}, 822 (1986), Yad. Fiz. {\bf 43}, 1282 (1986).

\bibitem{Sh1} Yu.M. Shabelski, Z. Phys. {\bf C38}, 569 (1988), doi:10.1007/BF01624362.

\bibitem{AMSHpA} G.H. Arakelyan, C. Merino, and Yu.M.~Shabelski, Eur. Phys. J. {\bf A52}, 9 (2016),  
doi:10.1140/epja/i2016-16009-2, e-Print:1509.05218[hep-ph].

\bibitem{Sha} Yu.M. Shabelski, Sov. J. Nucl. Phys. {\bf 50}, 149 (1989), Yad. Fiz. {\bf 50}, 239 (1989).

\bibitem{Shab} Yu.M. Shabelski, Z. Phys. {\bf C57}, 409 (1993), doi:10.1007/BF01474336.

\bibitem{JDDS} J. Dias de Deus and Yu.M. Shabelski, Phys. Atom. Nucl. {\bf 71}, 190 (2008),
doi:10.1134/S1063778808010213, Yad. Fiz. {\bf 71}, 191 (2008), e-Print:0612346[hep-ph].

\bibitem{AMSHpL} G.H.~Arakelyan, C.~Merino, and Yu.M.~Shabelski, Phys. Atom. Nucl. {\bf 77}, 626 (2014),
doi:10.1134/S1063778814050172, Yad. Fiz. {\bf 77}, 661 (2014), e-Print:1305.0388[hep-ph].

\bibitem{AnSh} V.V.~Anisovich and V.M.~Shekhter, Nucl. Phys. {\bf B55}, 455 (1973),
doi:10.1016/0550-3213(73)90391-X, Nucl. Phys. {\bf B63}, 542 (1973) (erratum),
doi:10.1016/0550-3213(73)90163-6.

\bibitem{CS} A. Capella and C.A. Salgado, Phys. Rev. {\bf C60}, 054906 (1999),
doi:10.1103/Phys\newline RevC.60.054906, e-Print:9903414[hep-ph].

\bibitem{AKMS} G.H. Arakelyan, A.B. Kaidalov, C. Merino, and Yu.M.~Shabelski, 
Phys. Atom. Nucl. {\bf 74}, 126 (2011), doi:10.1134/S1063778811030057, and e-Print:1004.4074[hep-ph].

\bibitem{Bocquet} G.~Bocquet {\it et al.}, UA1 Collaboration, Phys. Lett. {\bf B366}, 447 (1996),
doi:10.1016/\\0370-2693(95)01437-3.

\bibitem{SaTai} Sa Ben-Hao and Tai An, Phys. Lett. {\bf B399}, 29 (1997),
doi:10.1016/S0370-2693(97)\\00274-8, e-Print:9804003[nucl-th].

\bibitem{PACIAE} Sa Ben-Hao {\it et al.}, Comput. Phys. Commun. {\bf 184}, 1476 (2013),
doi:10.1016/j.cpc.\\2012.12.026, e-print:1206.4795[nucl-th].

\bibitem{PACIAE_1} She Zhi-Lei, {\it et al.}, Comput. Phys. Commun. {\bf 274}, 108289 (2022),
doi:10.1016/j.cpc.\\2022.108289.

\bibitem{Drescher} H.-J.~Drescher, J.~Aichelin, and K.~Werner, Phys. Rev. {\bf D65}, 057501 (2002),
doi:10.\\1103/PhysRevD.65.057501, e-Print:0105020[hep-ph].

\bibitem{Tounsi} A.~Tounsi and K.~Redlich, e-Print:0111159[hep-ph].

\bibitem{Tounsi1} J.S.~Hamieh, K.~Redlich, and A.~Tounsi, Phys. Lett. {\bf B486}, 61 (2000),
doi:10.1016/\\S0370-2693(00)00762-0, e-Print:0006024[hep-ph].

\bibitem{UrQMD} M.~Bleicher, W.~Greiner, H.~St{\"o}ker, and N.~Xu, Phys. Rev. {\bf C62}, 061901 (2000),
doi:10.1103/PhysRevC.62.061901, e-Print:0007215[hep-ph].

\bibitem{WA97} E.~Andersen {\it et al.}, WA97 Collaboration, Phys. Lett. {\bf B449}, 401 (1999),
doi:10.1016/\\S0370-2693(99)00140-9.

\bibitem{NA5722} N.~Carrer, NA57 and WA97 Collaborations, contribution to the Proceedings of
the 16th International Conference on Strangeness in Quark Matter (Quark Matter 200i), Berkeley, CA (USA),
20-25 July 2000, J. Phys. {\bf G27}, 391 (2001), doi:10.1088/0954-3899/27/3/317.

\bibitem{NA5222} S.~Kabana {\it et al.}, NA52 Collaboration, contribution to the Proceedings of
the 16th International Conference on Strangeness in Quark Matter (Quark Matter 200i), Berkeley, CA (USA),
20-25 July 2000, J. Phys. {\bf G27}, 495 (2001), doi:10.1088/0954-3899/27/3/329, e-Print:0010053[hep-ex].

\bibitem{Becattini} F.~Becattini {\it et al.}, Phys. Lett. {\bf B632}, 233 (2006),
doi:10.1016/j.physletb.2005.\newline10.053, e-Print:0508188[hep-ph].

\bibitem{Cleymans} J.~Cleynmans, B.~K{\"a}mpfer, and S.~Wheaton, contribution to the Proceedings of
the 16th International Conference on Ultra-Relativistic Nucleus-Nucleus Collisions
(Quark Matter 2002), Nantes (France), 18-24 July 2002, Nucl. Phys. {\bf A715}, 553 (2003),
doi:10.1016/S0375-9474(02)01517-8, e-Print:0208247[hep-ph].

\bibitem{Becattini1} F.~Becattini {\it et al.}, Phys. Rev. {\bf C69}, 024905 (2004),
doi:10.1103/PhysRevC.69.\newline024905, e-Print:0310049[hep-ph].

\bibitem{Fochler} O.~Fochler {\it et al.}, contribution to the Proceedings of the 43rd International Winter
Meeting in Nuclear Physics (Bormio 2005), Bormio (Italy), 13-20 March 2005, e-Print050502[hep-ph].

\bibitem{Lobanov} S.~Lobanov, A.~Maevskiy, and L.~Smirnova, contribution to the Proceedings of the International
Workshop on the March (IHEP-LHC-2011), Protvino (Russian Federation), 16-18 November 2011, PoS IHEP-LHC-2011,
{\bf 168}, 006 (2011), doi:10.22323/1.168.0008.

\bibitem{Orava} P.K.~Malhotra and R.~Orava, Z. Phys. {\bf C17}, 85 (1983),
doi:10.1007/BF01577823.

\bibitem{Wroblewski} A.~Wroblewski, Acta Phys. Pol. {\bf B16}, 379 (1985).

\bibitem{Long} H.-Y. Long {\it et al.}, Phys. Rev {\bf C84}, 0349 (2011),
doi:10.1103/PhysRevC.84.034905, e-Print:1103.2618[hep-ph].

\bibitem{ALICE22} K.~Aamodt {\it et al.}, ALICE Collaboration, Eur. Phys. J. {\bf C68}, 345 (2010),
doi:10.\\1140/epjc/s10052-010-1350-2, e-Print:1004.3514[hep-ex], and references therein.

\bibitem{CMS22} V.~Khachatryan {\it et al.}, CMS Collaboration, JHEP {\bf 01}, 079 (2011),
doi:10.1007/\\JHEP01(2011)079, e-Print:1011.5531[hep-ex], and references therein.

\bibitem{ATLAS22} G.~Aad {\it et al.}, ATLAS Collaboration, New J. Phys. {\bf 13}, 053033 (2011),
doi:10.1088/\\1367-2630/13/5/053033, e-Print:1012.5104[hep-ex], and references therein.

\bibitem{Satz} H.~Satz, EPJ Web of Conferences {\bf 171}, 02005 (2018),
doi:10.1051/epjconf/\newline201817102005.

\bibitem{AmelinBravina} N.S.~Amelin and L.V.~Bravina, Sov. J. Nucl. Phys {\bf 51}, 133 (1990), Yad. Fiz. {\bf 51}, 211 (1990).

\bibitem{AmelinBravina1} N.S.~Amelin {\it et al.}, Sov. J. Nucl. Phys {\bf 51}, 535 (1990), Yad. Fiz. {\bf 51}, 841 (1990).

\bibitem{LAQGSM} K.K.~Gudima, S.G.~Mashnik, and A.J.~Sierk, Los Alamos National Report LA-UR-01-6804 (2001).

\bibitem{CEM2k} S.G.~Mashnik, and A.J.~Sierk, Proceedings of AccApp00, La Grange Park, IL (USA), 328 (2001),
e-print:0011064[nucl-th].

\bibitem{Mashnik} S.G.~Mashnik {\it et al.}, Adv. Space Res. {\bf 34}, 1288 (2004), Proceedings of 34th COSPAR
Scientific Assembly: The 2nd World Space Congress, doi:10.1016/j.asr.2003.08.057, e-print:0210065[nucl-th]

\bibitem{QGSJETII} S.~Ostapchenko, Phys. Rev. {\bf D83}, 014018 (2011),
doi:10.1103/PhysRevD.83.014018, e-print:1010.1869[hep-ph].

\bibitem{Bleibel} J.~Bleibel, L.V.~Bravina, and E.E. Zabrodin, Phys. Rev. {\bf D93}, 114012 (2016),
doi:10.1103/PhysRevD.93.114012, e-print:1011.2703[hep-ph].

\bibitem{Amelin_PRC47} N.S.~Amelin {\it et al.}, Phys. Rev. {\bf C47}, 2299 (1993),
doi:10.1103/PhysRevC.47.2299. 

\bibitem{NA35} J.~Bartke {\it et al.}, NA35 Collaboration, Z. Phys. {\bf C48}, 191 (1990),
doi:10.1007/BF0155\\4465.

\bibitem{NA35_1} J.~B{\"a}chler {\it et al.}, NA35 Collaboration, Nucl. Phys. {\bf A525}, 221c (1991),
Proceedings of Quark Matter 90: VIIIth International Conference on Ultrarelativistic Nucleus-Nucleus Collisions (QM90),
Menton (France), 7-11 May 1990, doi:10.1007/BF01554465.

\bibitem{ABP} N.S. Amelin, M.A. Braun, and C. Pajares, Phys. Lett. {\bf B306}, 312 (1993),
doi:10.\\1016/0370-2693(93)90085-V.

\bibitem{ABPZP} N.S. Amelin, M.A. Braun, and C. Pajares, Z. Phys. {\bf  C63}, 507 (1994),
doi:10.1007/\\BF01580331.

\bibitem{AGK} V.A. Abramovsky, V.N. Gribov, and O.V.~Kancheli, Sov. J. Nucl. Phys. {\bf 18}, 308 (1973),
Yad. Fiz. {\bf 18}, 595 (1973).

\bibitem{TerMartirosyan_73} K.A.~Ter-Martirosyan, Phys. Lett. {\bf 44}, 179 (1973), doi:10.1016/0370-2693(73)90516-9;
Phys. Lett. {\bf 44}, 377 (1973), doi:10.1016/0370-2693(73)90411-5.

\bibitem{Gribov_1967} V.N.~Gribov, Sov. Phys. JETP {\bf 26}, 414 (1968), Zh. \'Eksp. Teor. Fiz. {\bf 53}, 654 (1967).

\bibitem{Kai} A.B. Kaidalov, Sov. J. Nucl. Phys. {\bf 45}, 902 (1987), Yad. Fiz. {\bf 45}, 1452 (1987).

\bibitem{BT} L. Bertocchi and D. Treleani, J. Phys. {\bf G3}, 147 (1977), doi:10.1088/0305-4616/3/2/007.

\bibitem{Weis} J. Weis, Acta Phys. Polon. {\bf B7}, 851 (1976).

\bibitem{Sh3} Yu.M. Shabelski, Sov. J. Nucl. Phys. {\bf 26}, 573 (1977), Yad. Fiz. {\bf 26}, 1084 (1977),
Nucl. Phys. {\bf B132}, 491 (1978), doi:10.1016/0550-3213(78)90473-X.

\bibitem{Jar} T. Jaroszewicz, J.~Kwiecinski, L.~Lesniak, and K.~Zalewski, Z. Phys. {\bf C1}, 181 (1979),
doi:10.1007/BF01445409.

\bibitem{Alk} G.D. Alkhazov {\it et al.}, Nucl. Phys. {\bf A280}, 365 (1977),
doi:10.1016/0375-9474(77)\\90611-X.

\bibitem{Phob} B.B. Back {\it et al.}, PHOBOS Collaboration, Phys. Rev. Lett. {\bf 85}, 3100 (2000),
doi:10.\\1103/PhysRevLett.85.3100, e-Print:0007036[hep-ex].

\bibitem{Phen} K. Adcox {\it et al.}, PHENIX Collaboration, Phys. Rev. Lett. {\bf 86}, 3500 (2001),
doi:10.\\1103/PhysRevLett.86.3500, e-Print:0012008[nucl-ex].

\bibitem{CMT} A. Capella, C. Merino, and J. Tran Thanh Van, Phys. Lett. {\bf B265}, 415 (1991),
Nucl. Phys {\bf A544}, 581 (1992), doi:10.1103/PhysRevD.45.92.

\bibitem{AP} N. Armesto and C. Pajares, Int. J. Mod. Phys. {\bf A15}, 2019 (2000),
doi:10.1142/\\S0217751X00000823, e-Print:0002163[hep-ph].

\bibitem{Schw} A.~Schwimmer, Nucl. Phys. {\bf B94}, 445 (1975), doi:10.1016/0550-3213(75)90106-6.

\bibitem{ABFP} N. Armesto, M.A. Braun, E.G. Ferreiro, and C. Pajares, Phys. Rev. Lett. {\bf 77}, 3736 (1996),
doi:10.1103/PhysRevLett.77.3736, e-Print:9607239[hep-ph]. 

\bibitem{AMSHnphi} G.H.~Arakelyan, C.~Merino, and Yu.M.~Shabelski, Int. J. Mod. Phys. {\bf A33}, 1850202 (2018),
doi:10.1142/S0217751X18502020, e-Print:1810.02170[hep-ph]. 

\bibitem{AMSHKst} G.H.~Arakelyan, C.~Merino, and Yu.M.~Shabelski, Eur. Phys. J. {\bf A55}, 151 (2019), 
doi:10.1140/epja/i2019-12832-1, e-Print:1805.11419[hep-ph].

\bibitem{NA49pbp} T.~Anticic {\it et al.}, NA49 Collaboration, Phys. Rev. {\bf C69}, 024902 (2004),
doi:10.1103/\\PhysRevC.69.024902.

\bibitem{NA49c} T.~Anticic {\it et al.}, NA49 Collaboration, Phys. Rev. Lett. {\bf 93}, 022302 (2004),
doi:10.\\1103/PhysRevLett.93.022302, e-Print:0311024[nucl-ex].

\bibitem{NA49d} C.~Alt {\it et al.}, NA49 Collaboration, Phys. Rev. Lett. {\bf 94}, 192301 (2005),
doi:10.1103/\\PhysRevLett.94.192301, e-Print:0409004[nucl-ex].

\bibitem{NA49b} C.~Alt {\it et al.}, NA49 Collaboration, Phys. Rev. {\bf C78}, 034918 (2008),
doi:10.1103/\\PhysRevC.78.034918, e-Print:0804.3770[nucl-ex].

\bibitem{NA49a} T. Anticic {\it et al.}, NA49 Collaboration, Phys. Rev. {\bf C80}, 034906 (2009),
doi:10.1103/\\PhysRevC.80.034906, e-Print:0906.0469[nucl-ex].

\bibitem{BM} C. Blume and C. Markert, Prog. Part. Nucl. Phys. {\bf 66}, 834 (2011),
doi:10.1016/\\j.ppnp.2011.05.001, e-Print:1105.2798[nucl.ex]. 

\bibitem{NA57b} F. Antinori {\it et al.}, NA57 Collaboration, Phys. Lett. {\bf B595}, 68 (2004),
doi:10.1016/\\j.physletb.2004.05.025, e-Print:0403022[nucl-ex].

\bibitem{NA57} F. Antinori {\it et al.}, NA57 Collaboration, J. Phys. {\bf G32}, 427 (2006),
doi:10.1088/0954-3899/32/4/003, e-Print:0601021[nucl-ex].

\bibitem{STAR3} J. Adams {\it et al.}, STAR Collaboration, Phys. Rev. Lett. {\bf 92}, 182301 (2004),
doi:10.\\1103/PhysRevLett.92.182301, e-Print:0307024[nucl-ex].

\bibitem{STAR2} J. Adams {\it et al.}, STAR Collaboration, Phys. Rev. Lett. {\bf 98}, 062301 (2007),
doi:10.\\1103/PhysRevLett.98.062301, e-Print:0606014[nucl-ex].

\bibitem{STARp} B.I. Abelev {\it et al.}, STAR Collaboration, Phys. Rev.{\bf C79}, 034909 (2009),
doi:10.\\1103/PhysRevC.79.034909, e-Print:0808.2041[nucl-ex].

\bibitem{STAR} M.M.~Aggarwal {\it et al.}, STAR Collaboration, Phys. Rev. {\bf C83}, 024901 (2011),
doi:10.\\1103/PhysRevC.83.024901, e-Print:1010.0142[nucl-ex].

\bibitem{STAR1} G. Agakishiev {\it et al.}, STAR Collaboration, Phys. Rev. Lett. {\bf 108}, 072301 (2012),\\
doi:10.1103/PhysRevLett.108.072301, e-Print:1107.2955[nucl-ex].

\bibitem{PHENIX} K.~Adcox {\it et al.}, PHENIX Collaboration, Phys. Rev. Lett.{\bf 89}, 092302 (2002),
doi:10.\\1103/PhysRevLett.89.092302, e-Prinr:0204007[nucl-ex].

\bibitem{PHENIXp} K. Adcox {\it et al.}, PHENIX Collaboration, Phys. Rev. {\bf C69}, 024904 (2004),
doi:10.\\1103/PhysRevC.69.024904, e-Print:0307010[nucl-ex].

\bibitem{ALICEPbp276} B. Abelev {\it et al.}, ALICE Collaboration, Phys. Rev. {\bf C88}, 044910 (2013),
doi:10.\\1103/PhysRevC.88.044910, e-Print:1303.0737[hep-ex].

\bibitem{ALICEms276} B. Abelev {\it et al.}, ALICE Collaboration, Phys. Lett. {\bf B728}, 216 (2014),
doi:10.1016/\\j.physletb.2013.11.048,
Phys. Lett. {\bf B734}, 409 (2014) (erratum), doi:10.1016/\\j.physletb.2014.05.052, e-Print:1307.5543[nucl-ex].

\bibitem{ALICEPbp5} S.~Acharya {\it et al.}, ALICE Collaboration, Phys. Rev. {\bf C101}, 044907 (2020),
doi:10.\\1103/PhysRevC.101.044907, e-Print:1910.07678[nucl-ex].

\end{thebibliography}
\end{document}